\documentclass{article}
\usepackage{arxiv}
\usepackage[T1]{fontenc}
\usepackage[utf8]{inputenc} 
\usepackage{amsmath}
\usepackage{amsfonts}

\usepackage{graphicx}
\graphicspath{{./images/}}
\usepackage{subcaption}

\usepackage{booktabs}
\usepackage{array}
\usepackage{multirow}
\usepackage[table]{xcolor}
\definecolor{lightgray}{gray}{0.85}

\usepackage{nicefrac}
\usepackage{microtype}
\usepackage{url}
\usepackage{hyperref}
\usepackage{lipsum}

\title{Design of a specimen to train path-dependent deep learning material models from a single uniaxial test: eliciting strain diversity via automatically differentiable elastoplastic topology optimization}

\author{
    Shunyu Yin \\
    School of Engineering\\
    Brown University\\
    184 Hope St, Providence, RI 02912, USA\\
    \And
    Bernardo P. Ferreira\\
    School of Engineering\\
    Brown University\\
    184 Hope St, Providence, RI 02912, USA\\
    \And
    Gawe\l{} Ku\'s\\
    School of Engineering\\
    Brown University\\
    184 Hope St, Providence, RI 02912, USA\\
    \And
    Miguel A. Bessa\\
    School of Engineering\\
    Brown University\\
    184 Hope St, Providence, RI 02912, USA\\
    \texttt{miguel\_bessa@brown.edu}\\
}

\begin{document}
\maketitle
\begin{abstract}
Artificial neural networks accurately learn nonlinear, path-dependent material behavior. However, training them typically requires large, diverse datasets, often created via synthetic unit cell simulations. This hinders practical adoption because physical experiments on standardized specimens with simple geometries fail to generate sufficiently diverse stress–strain trajectories. Consequently, an unreasonably large number of experiments or complex multi-axial tests would be needed. This work shows that such networks can be trained from a single specimen subjected to simple uniaxial loading, by designing the specimen using a novel automatically differentiable elastoplastic topology optimization method. Our strategy diversifies the stress–strain states observed in a single test involving plastic deformation. We then employ the automatically differentiable model updating (ADiMU) method to train the neural network surrogates. This work demonstrates that topology-optimized specimens under simple loading can train large neural networks, thereby substantially reducing the experimental burden associated with data-driven material modeling.
\end{abstract}


\section{Introduction}
\label{sec:intro}

Materials underpin most scientific and technological innovations. For centuries, we have understood, modeled, and experimentally characterized them based on constitutive laws \cite{lemaitre1994mechanics, simo1998computational, malvern1969introduction}. These laws predict linear and nonlinear behavior, from mechanical to chemical and electromagnetic phenomena. Material elasticity is a quintessential example, being described by constitutive models with a small number of parameters, determined from experiments of standardized specimens with simple geometries. However, materials are also path- or history-dependent \cite{ hill1998mathematical,truesdell2004non,de2008computational}. They can remember, evolve, permanently deform, fatigue, and fail -- phenomena emerging from complex matter interactions. Sometimes this can be captured from theories grounded in first-principles, but often it needs phenomenological or empirical reasoning. Until recently, material models were developed to have as few parameters as possible, such that they could be calibrated with a small number of experiments. In contrast, machine learning models have been shown to effectively capture complex material behavior \cite{ghaboussi1998autoprogressive,bessa:2017a,mozaffar_deep_2019}, but contain a large number of parameters. Therefore, these models have required large datasets that so far were only obtained from \textit{synthetic} material experiments \cite{ferreira_automatically_2025}, i.e., computational simulations instead of physical experiments.

Despite the need to train machine learning–based constitutive models from computational simulations of materials, these models are experiencing a surge in the literature due to their accuracy and efficiency trade-off \cite{zhang2022learning, linden2023neural}. They replace phenomenological material models by establishing a map (regression) between material deformations or strains (a six-dimensional input vector representing a 3$\times$3 symmetric tensor) and stresses (a six-dimensional output vector representing a 3$\times$3 symmetric tensor). These nonlinear maps can also be non-injective due to history- or path-dependency in materials (a stress-strain state of a material depends on the complete deformation path up to that state). Deep neural networks with recurrent units such as recurrent neural networks (RNNs) can effectively model this behavior, as first shown in \cite{mozaffar_deep_2019} and later extended by others, e.g., \cite{wang_general_2020, gorji_potential_2020, maia_physically_2023}. Subsequent studies focused on improving training efficiency and incorporating physics constraints to reduce the amount of data needed to train these models \cite{borkowski_recurrent_2022, eghbalian_physics-informed_2023, wang_plastic_2024}, increasing their applicability. Nevertheless, neural network material models still need to be trained with diverse datasets that explore a large number of stress-strain states and paths \cite{ferreira_automatically_2025}.

\begin{figure}[h!]
	\centering
	\includegraphics[width=0.70\textwidth]{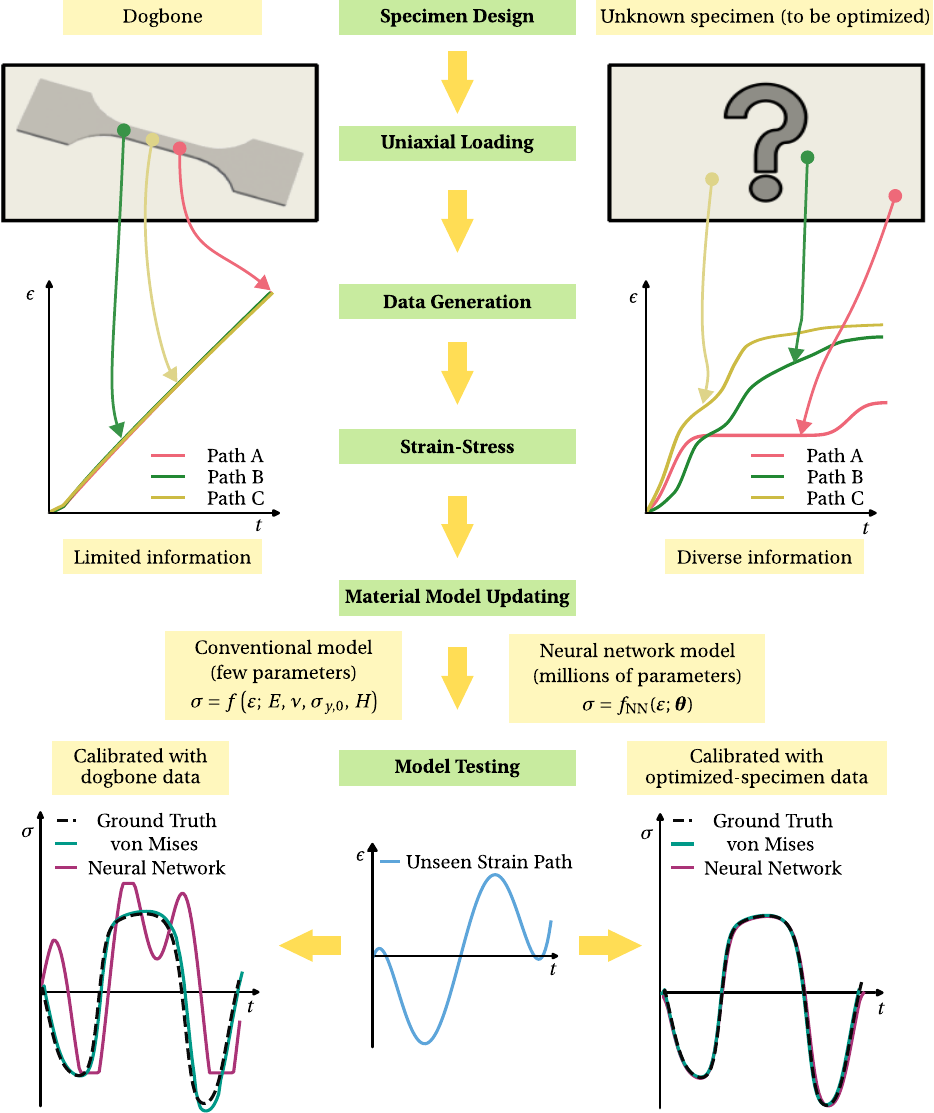}
	\caption{What is the optimal geometry to train (update or calibrate) a material model? The standard dogbone specimen (left) generates limited stress–strain paths under uniaxial loading, leading to less informative data that can only train a conventional material model with few parameters; the dogbone specimen is inadequate to train a neural network material model with millions of parameters. In contrast, the aim of this article is to find an optimized specimen geometry (right) that generates diverse strain–stress data from a single uniaxial test from which a neural network material model can be trained.}
	\label{fig:motivation}
\end{figure}

Fig. \ref{fig:motivation} summarizes the importance of stress-strain diversity. The left panel shows that a conventional model (von Mises plasticity) can be successfully trained using limited data from a standard dogbone test \cite{ferreira_automatically_2025}. Its prediction (solid green line) matches the ground-truth (dashed line) for an unseen strain path. However, the same dataset fails to train a neural network model (purple line), which cannot accurately predict the response. The network's high parameter count requires a richer dataset than a simple dogbone test can provide. Conversely, the right panel asks: can we design a \textbf{single} specimen under \textbf{uniaxial loading} that generates sufficient strain diversity to train a path-dependent neural network?

This gap is addressed for the first time in this article by performing topology optimization of a specimen undergoing plastic deformation, followed by material model updating. We aim to create a complex geometry that generates diverse deformation states under uniaxial load. Such geometries are specifically intended to enable, in future experimental studies, the effective use of full-field measurement techniques (e.g., DIC, DVC) to capture rich strain information \cite{bigger_good_2018, leclerc2009integrated, denys2016multi, ricciardi_advancements_2025}. Therefore, we support the ``Material Testing 2.0'' paradigm \cite{pierron_material_2023} by replacing standardized tests with a single, information-rich experiment. Such one-shot identification has been demonstrated for hyperelasticity \cite{guelon_new_2009} and anisotropic yield \cite{fu_method_2020}. Here, we employ the Automatically Differentiable Model Updating (ADiMU) framework and its implementation, HookeAI \cite{ferreira_automatically_2025}\footnote{\url{https://github.com/bessagroup/hookeai}}. While model updating has a long history \cite{ghaboussi1998autoprogressive,kavanagh1971finite,grediac:1989a, shin:2000a,lefik:2003a,mathieu_estimation_2015,gerbig2016coupling,thakolkaran:2022a, wu:2025a, akerson:2025a,chen_finite_2025}, our approach integrates optimization with deep learning to maximize data efficiency. Our work contrasts with early studies that used parametric shape optimization (e.g., adding holes or notches) to vary strain states \cite{demmerle_optimal_1993, creuziger_insights_2017}. These works were done by either minimizing parameter uncertainty through coupled material model calibration \cite{bertin_optimization_2016, chapelier_spline-based_2022} or maximizing strain diversity independently \cite{souto_numerical_2016, andrade2019design,ihuaenyi_seeking_2024, tung_anti-dogbone_2024, ihuaenyi_mechanics_2025, ricciardi_bayesian_2024}. While these modifications enhance diversity, they rely on trial-and-error or low-dimensional design spaces, limiting their effectiveness. More recently, researchers have applied topology optimization to design complex specimens by distributing material within a discretized domain \cite{bendsoe_optimal_1989, bendsoe_topology_2004, chamoin_coupling_2020}. Notable examples include density-based designs inspired by compliant mechanisms \cite{barroqueiro_design_2020, gonccalves2022topology, goncalves_design_2023} and frameworks for calibrating linear elastic parameters \cite{ghouli_topology_2025}. Although various specimens were designed in the aforementioned contributions, whether through shape optimization or topology optimization, the evaluations of these designs either focused solely on demonstrating strain/stress-field heterogeneity based on their chosen metrics or were limited to calibrating simple linear elastic material parameters, where simple tests often suffice.

We note that designing specimens via elastoplastic topology optimization remains largely unexplored, likely due to the computational challenges of plasticity and the difficulty of defining metrics for strain diversity. In particular, optimizing for diversity in plastic deformation requires a differentiable objective function. Conventional elastoplastic topology optimization relies on manually derived adjoint formulations \cite{yuge_optimization_1995, maute_adaptive_1998, amir_stress-constrained_2017, alberdi_unified_2018}, which can be laborious and restrict the exploration of complex objective functions. Automatic differentiation (AD) offers a powerful alternative \cite{margossian_review_2019}, enabling exact gradient computation by unrolling time-stepping integration in plasticity simulation \cite{ferreira_automatically_2025,jia_multimaterial_2025,jia_multimaterial_2025-1}.

\begin{figure}[h!]
	\centering
	\includegraphics[width=0.70\textwidth]{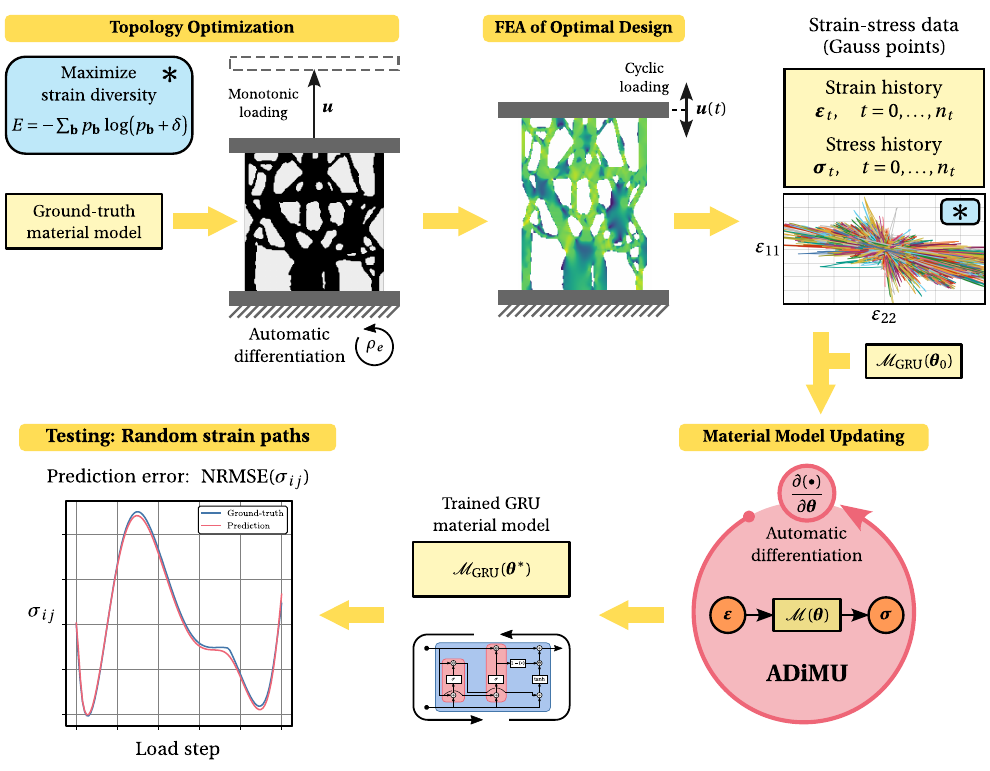}
	\caption{Schematic workflow of the proposed framework. Step 1: Automatically-differentiable elastoplastic topology optimization to find a specimen under uniaxial monotonic loading that maximizes the strain diversity within that specimen. Step 2: The optimized specimen is subjected to a cyclic tension–compression loading to generate a local stress–strain dataset that spans a wide range of deformation states, as illustrated by the cloud of trajectories in the $ \epsilon_{11} - \epsilon_{22}$ plane. Step 3: The effectiveness of the topology-optimized specimen is evaluated by training a GRU-based recurrent neural network model based on the dataset and testing on an unseen randomly generated polynomial dataset.}
	\label{fig:workflow}
\end{figure}

In summary, we leverage JAX-based AD and develop an elastoplastic topology optimization strategy to design specimens that maximize strain diversity under uniaxial tension. Rather than focusing on the calibration of traditional constitutive models, the optimized specimen is treated as a data generator to enable fully data-driven training of recurrent neural network surrogates based on Gated Recurrent Units (GRUs), using stress–strain data obtained from a single test. To the best of our knowledge, this is the first integration of elastoplastic
topology optimization with data-driven surrogate model training from a single optimized specimen. The overall workflow of the proposed framework is illustrated in Fig. \ref{fig:workflow}.

\section{Methodology}

The proposed framework integrates automatically differentiable topology optimization with data-driven training of
neural network material models. It consists of three main stages:

\begin{itemize}
    \item \textbf{Stage 1: Topology optimization.} Solves an elastoplastic topology optimization problem using AD and considering monotonic loading. Strain diversity is maximized according to an entropy-based objective (Section \ref{sec:to_framework}).
    \item \textbf{Stage 2: Dataset generation by finite element analysis.} Subjects the optimized specimen to cyclic loading to generate a local stress–strain dataset (Section \ref{data_gen}).
    \item \textbf{Stage 3: Material model updating (training).} Train a material model with the dataset, and test it on independent data to evaluate generalization. In this article, we will consider a large GRU-based neural network, as it is the most challenging model given its large parameter count (Section \ref{mat_model_updating}). Any other model could be trained instead.
\end{itemize}


\subsection{Topology optimization}
\label{sec:to_framework}

We optimize material distribution to maximize strain diversity. Adopting a standard density-based formulation \cite{bendsoe_topology_2004}, we discretize the domain into finite elements, assigning each a density $\rho_e \in [0, 1]$ (0=void, 1=solid), with intermediate values penalized to drive the solution toward a binary distribution.

We consider a thin sheet whose thickness is much smaller than its in-plane dimensions. Accordingly, the problem is analyzed both (i) under a two-dimensional plane-stress assumption and (ii) in a fully three-dimensional setting. The material is assumed to follow a von Mises elastoplastic constitutive model with isotropic hardening. During the topology optimization process, a modified SIMP (Solid Isotropic Material with Penalization) scheme is employed, with the penalty factor gradually increased according to a prescribed continuation strategy. The optimization objective is to maximize the spatial spread of strain components, based on the assumption that a broader distribution in strain space yields more diverse strain paths. To make this objective suitable for gradient-based optimization, we formulate it as the Shannon entropy of a differentiable strain-histogram, as detailed in the following section. Additional details of the optimization setup and parameter choices are provided in Appendix \ref{app: to_setting}. We build the topology optimization formulation as,
\begin{equation}
\begin{aligned}
\min_{\boldsymbol{\rho}} \quad
& J(\boldsymbol{\rho}) = -\,E(\boldsymbol{\rho}) \\[4pt]
\text{s.t.} \quad
& \mathbf{R}^n\!\big(\mathbf{u}^n(\hat{\boldsymbol{\rho}}),
\hat{\boldsymbol{\rho}}\big) = \mathbf{0},
\qquad n = 1, \dots, N_t, \\[6pt]
& \underline{\rho}
\;\le\;
\frac{1}{N_e}\sum_{e=1}^{N_e} \hat{\rho}_e
\;\le\;
\overline{\rho}, \\[6pt]
& 0 \le \rho_{\min} \le \rho_e \le 1,
\qquad e = 1, \dots, N_e .
\end{aligned}
\end{equation}
Here, $E(\boldsymbol{\rho})$ denotes the Shannon entropy of the strain distribution, whose mathematical form is introduced in the following section. $\mathbf{R}^n(\cdot)=\mathbf{0}$ denotes the discrete residual equations at load step $n$.

The entire topology optimization framework is implemented in JAX \cite{jax2018github}, leveraging its just-in-time compilation and fast automatic differentiation capabilities to obtain the sensitivities (gradient of the objective with respect to the density field). The design variables (element densities) are updated at each iteration using the Method of Moving Asymptotes (MMA) \cite{svanberg1987method}, a common gradient-based optimizer in topology optimization which provides robust convergence under the proposed objective and volume fraction constraints.

\subsubsection{Objective function: promoting strain diversity}

\begin{figure}[h]
  \centering
  \begin{subfigure}[b]{0.48\textwidth}
    \centering
    \includegraphics[width=\textwidth]{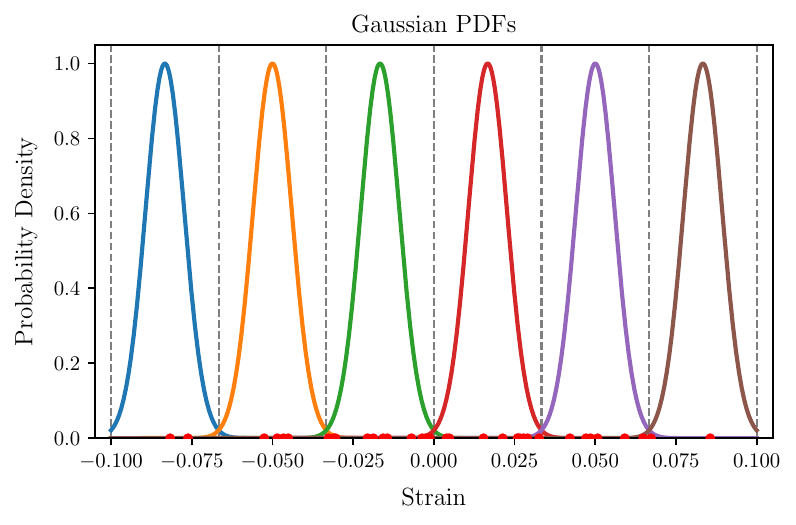}
    \caption{}
    \label{fig:gaussian}
  \end{subfigure}%
  \hfill
  \begin{subfigure}[b]{0.48\textwidth}
    \centering
    \includegraphics[width=\textwidth]{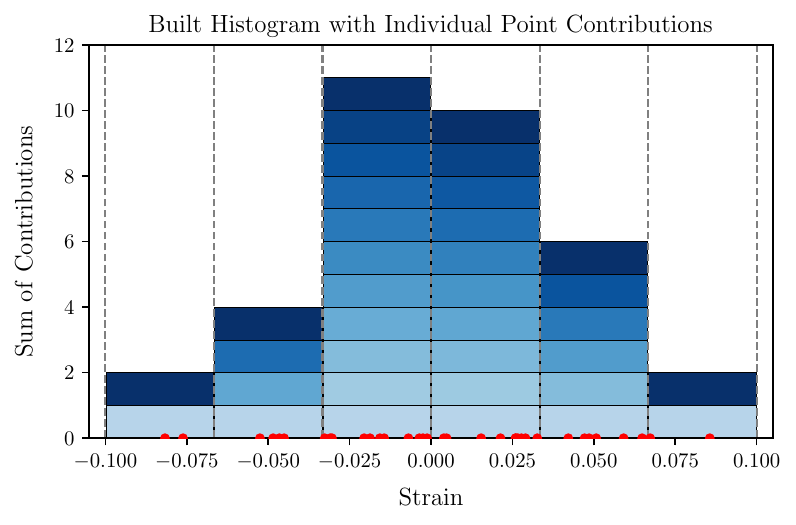}
    \caption{}
    \label{fig:hist}
  \end{subfigure}%
  \caption{One-dimensional schematic of the objective-function definition. (a) The predefined strain interval is partitioned into several equal-width cells. A Gaussian probability-density function (PDF), centered at the midpoint of each cell, is assigned as a weighting kernel. The standard deviation is selected so that the PDF decays to $\approx 0 $ outside its own cell; after normalization, a single strain sample therefore contributes $\approx 1 $ to its host cell and negligibly elsewhere. (b) An aggregate histogram is built by summing these individual contributions over all samples. Because each sample behaves as a unit impulse within its cell, the resulting bar heights closely approximate the discrete point counts in the corresponding bins, providing a differentiable surrogate for cell occupancy. The dataset achieves maximum diversity when the histogram is a uniform distribution.}
  \label{fig:1D}
\end{figure}

We introduce a differentiable, entropy-based objective function to quantify strain diversity. Diversity is defined as the coverage of strain states within a predefined space. 
To evaluate this coverage, the strain space is discretized into cells and populated by strain samples obtained from the design. Since discrete cell counting is non-differentiable, we adopt a soft assignment approach: each data point influences neighboring cells via a Gaussian kernel, enabling gradient-based optimization. This constructs a differentiable histogram of strain states. This idea is illustrated for one single strain component in Fig. \ref{fig:1D}, although we generalize it to all strain components.

Finally, the Shannon entropy of this histogram is computed, providing a differentiable scalar measure of strain diversity. Maximizing this entropy-based metric during optimization guarantees that the resulting specimen provides the broadest possible coverage of strain states. The step-by-step definition of this objective function is detailed as follows:

\begin{enumerate}
    \item \textbf{Discretization:} We define the strain space boundaries for each component and uniformly discretize each range into $n_b$ intervals, yielding $n_b^{n_s}$ total cells ($n_s$ is the number of strain components).

    \item \textbf{Soft Assignment:} To ensure differentiability, we assign a Gaussian PDF centered at each cell midpoint, allowing data points to continuously influence neighboring cells. For each strain point $\boldsymbol{\varepsilon}_{p}$ (final state under monotonic load) and cell center $\boldsymbol{c}_{b}$, we calculate a scaled distance for component $i$:
    \begin{equation}
    \text{D}_{p,b,i} = \frac{\varepsilon_{p,i} - c_{b,i}}{\Delta_i \cdot s}\,,
    \end{equation}
    where \(\Delta_i\) is the cell width and \( s = \frac{1}{6} \) is a scaling factor chosen to create narrow kernels that emulate hard cell counting while maintaining differentiability. The Gaussian kernel is:
    \begin{equation}
    \varphi_{p,b,i} = \frac{1}{\sqrt{2\pi}} \exp\left(-\frac{1}{2}\text{D}_{p,b,i}^2\right)\,,
    \end{equation}

    \item \textbf{Joint PDF:} We compute a joint PDF for each data point by multiplying component contributions:
    \begin{equation}
    \Phi_{p,\mathbf{b}} = \prod_{i=1}^{n_s}\varphi_{p,b,i}\,,
    \end{equation}
    where $ \mathbf{b} $ is the tuple of cell indices.

    \item \textbf{Normalization:} Joint contributions are normalized so each data point sums to unity:
    \begin{equation}
    \Bar{\Phi}_{p,\mathbf{b}} = \frac{\Phi_{p,\mathbf{b}}}{\sum_{\mathbf{b}} \Phi_{p,\mathbf{b}}}\,,
    \end{equation}

    \item \textbf{Histogram Construction:} We scale contributions by element density ($\rho^\mathrm{SIMP}$) so solid elements contribute more. Summing these yields a differentiable strain histogram:
    \begin{equation}
    H_{\mathbf{b}} = \sum_{p} \rho^{\mathrm{SIMP}}_p \, \Bar{\Phi}_{p,\mathbf{b}}\,,
    \end{equation}
    This is converted to a probability distribution:
    \begin{equation}
    p_{\mathbf{b}} = \frac{H_{\mathbf{b}}}{\sum_{\mathbf{b}} H_{\mathbf{b}}}\,,
    \end{equation}
    
    \item \textbf{Entropy Calculation:} Finally, we compute the Shannon entropy:
    \begin{equation}
    E = -\sum_{\mathbf{b}} p_{\mathbf{b}} \log(p_{\mathbf{b}}+\delta)\,.
    \label{eq:entropy}
    \end{equation}
    where $\delta$ ensures numerical stability.
\end{enumerate}

 The optimization framework aims to maximize the entropy $E$ of the resulting histogram (minimize the negative entropy $-E$), thereby achieving maximum spatial diversity of the strain dataset.

\subsubsection{Postprocessing}

We employ a standard cone filter \cite{bourdin_filters_2001} and threshold projection \cite{xu_volume_2010} (Appendix \ref{app:cone_filter}) in the topology optimization framework. To ensure a strictly black-and-white design, we apply a final post-processing step \cite{sigmund_benchmarking_2022} where densities are sorted and the top $n_f$ elements (satisfying the volume fraction) are set to 1, while others are set to 0.001,
\begin{equation}
v_f = \frac{1}{n_e}\sum_{e=1}^{n_e} \rho_e \,,\quad
n_f = \frac{{n_e} (v_f - 0.001)}{1 - 0.001},
\label{eq:threshold}
\end{equation}
\noindent where $n_e$ is the total number of elements. A binary morphological cleaning then removes isolated voids or material islands smaller than 20 pixels.

\subsection{Dataset generation by finite element analysis}
\label{data_gen}
The design that results from the automatically differentiable elastoplastic topology optimization process, is then subjected to cyclic uniaxial loading via finite element analysis (FEA). Local stress-strain paths from all solid elements are collected into a dataset that will then be employed to train a history-dependent material model.

\subsection{Material model updating: training a recurrent neural network}
\label{mat_model_updating}

The dataset is then used by the previously mentioned model updating method -- ADiMU \cite{ferreira_automatically_2025} via its HookeAI implementation -- to train a recurrent neural network material model. Note that this mimics the real scenario of finding a model when the material model is not known (in this case, we generate data using von Mises plasticity).

Importantly, we decided to train an overparameterized RNN material model containing GRU cells \cite{chung_gated_2015} to test if our single-specimen dataset is sufficiently diverse. This follows from the establishment of RNNs with GRUs as general material models capable of learning history-dependent mechanical behavior from diverse stress-strain datasets \cite{mozaffar_deep_2019}. Our model consists of two GRU layers (500 hidden units each) followed by a linear layer, totaling over two million parameters (implemented in PyTorch). Using such a flexible, overparameterized model serves as a rigorous test; if successful, training more constrained or physics-informed architectures \cite{borkowski_recurrent_2022, eghbalian_physics-informed_2023, wang_plastic_2024} that have significantly fewer parameters is also possible.

Data is standardized (zero mean, unit variance) and sequences of deformation states are 200 time steps long. The dataset is split as 80/10/10 (training/validation/testing). We minimize the Mean Squared Error (MSE) using Adam (LR=0.001 with exponential decay) for up to 200 epochs (batch size 32), with early stopping based on validation loss.

\subsubsection{Evaluation and testing procedure}

\begin{figure}[h!]
  \centering
  \begin{subfigure}[b]{0.48\textwidth}
    \centering
    \includegraphics[width=\textwidth]{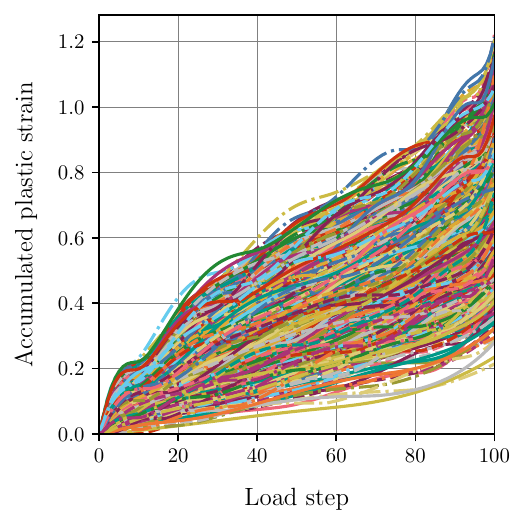}
    \caption{}
    \label{fig:ep_bar_plot_2D}
  \end{subfigure}%
  \hfill
  \begin{subfigure}[b]{0.48\textwidth}
    \centering
    \includegraphics[width=\textwidth]{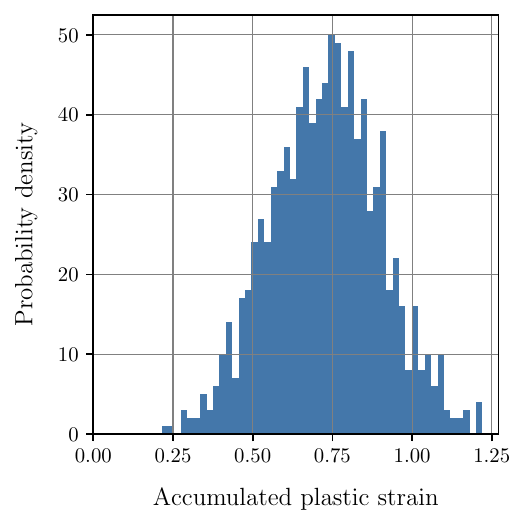}
    \caption{}
    \label{fig:ep_bar_histogram_2D}
  \end{subfigure}
  \hfill
  \begin{subfigure}[b]{0.48\textwidth}
    \centering
    \includegraphics[width=\textwidth]{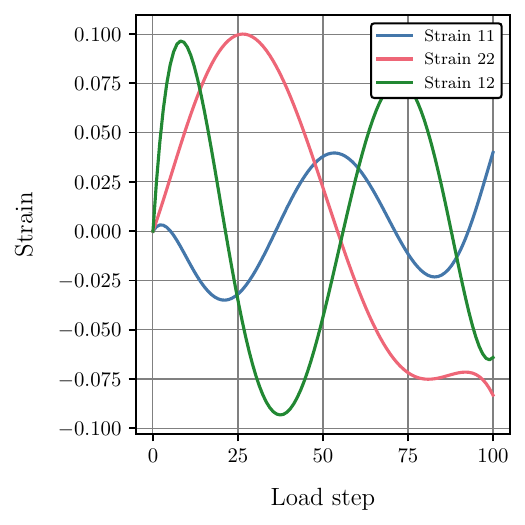}
    \caption{}
    \label{fig:strain_path_sample_2D}
  \end{subfigure}
  \hfill
  \begin{subfigure}[b]{0.48\textwidth}
    \centering
    \includegraphics[width=\textwidth]{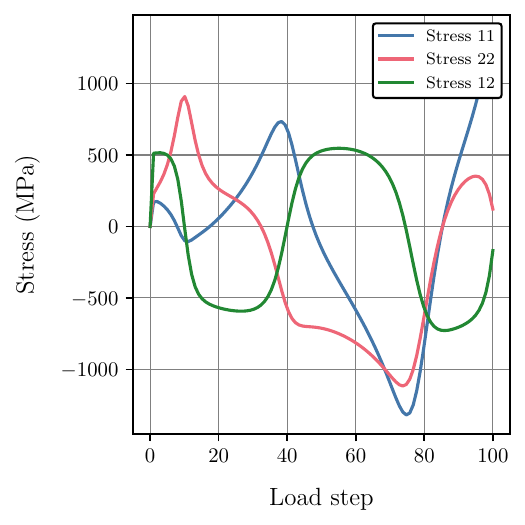}
    \caption{}
    \label{fig:stress_path_sample_2D}
  \end{subfigure}
  \caption{Visualization of the 2D test set: 
    (a) Accumulated plastic strain versus load step, demonstrating that all test paths undergo plastic deformation; 
    (b) Probability density distribution of the accumulated plastic strain across the test set; 
    (c) A representative sample of strain paths; 
    (d) The corresponding stress response for the strain paths shown in (c).}

  \label{fig:2D_testset_sample}
\end{figure}

In order to test the GRU's performance when predicting a wide variety of loading scenarios, a randomly generated polynomial testing set is created, as proposed in \cite{mozaffar_deep_2019} and implemented in HookeAI. A sequence of $100$ load increments is defined, and random strain values within predefined bounds are assigned to several control points along this sequence. The strain paths for each component are generated independently by connecting the control points using linear regression with polynomial basis functions as an interpolator.  Each strain component is sampled within $[-10\%, 10\%]$. Within these bounds, all generated testing paths enter the plastic regime and accumulate sufficient plastic strain for a meaningful evaluation of the GRU models, as shown in Fig. \ref{fig:ep_bar_plot_2D}.  An example trajectory from the 2D test set is presented in Fig. \ref{fig:strain_path_sample_2D}. The generation of the 3D test set is described in Appendix \ref{app:3D_testset}.

\section{Results}

In this section, we first present the optimized specimen geometry obtained from the proposed topology optimization framework. This specimen is then employed as a data generator to produce an optimized-specimen stress–strain dataset. Based on this dataset, GRU models are trained and tested within the ADiMU framework. This process not only evaluates the effectiveness of the optimization methodology but also creates a neural network surrogate model capable of capturing the elastoplastic constitutive behavior.

\subsection{Topology-Optimized Specimen}

The design domain uses a $150 \times 150$ mesh with a density filter radius of $r=4$ (see Appendices \ref{app:filter} and \ref{app:mesh_dependent} for parameter studies). We define the strain space using bounds from Tab. \ref{tab:strain_space}, discretized into $n_b = 10$ intervals per component, creating $1000$ cells for the optimization to target.

\begin{table}[h!]
\caption{Predefined strain‐space bounds used for objective calculation.}
\label{tab:strain_space}
\centering
\setlength{\tabcolsep}{0.30cm}
\renewcommand{\arraystretch}{1.5}
\newcolumntype{C}[1]{>{\centering\arraybackslash}p{#1}}
\begin{tabular}{C{3cm} C{3cm} C{3cm}}
\toprule
Strain 11       & Strain 22       & Strain 12       \\
\midrule
$[-0.1,\,0.1]$  & $[0.0,\,0.1]$   & $[-0.1,\,0.1]$  \\
\bottomrule
\end{tabular}

\vspace{0.2em}
\small \textit{Note: Strain 22 corresponds to the loading direction.}
\end{table}

The optimized design undergoes thresholding and morphological cleaning to remove small features (Fig. \ref{fig:three_designs}). This final geometry generates the datasets for valid GRU training. 3D results are in Appendix \ref{app:3D_results}.

\begin{figure}[h!]
  \centering
  \begin{subfigure}[b]{0.28\textwidth}
    \centering
    \includegraphics[width=\textwidth]{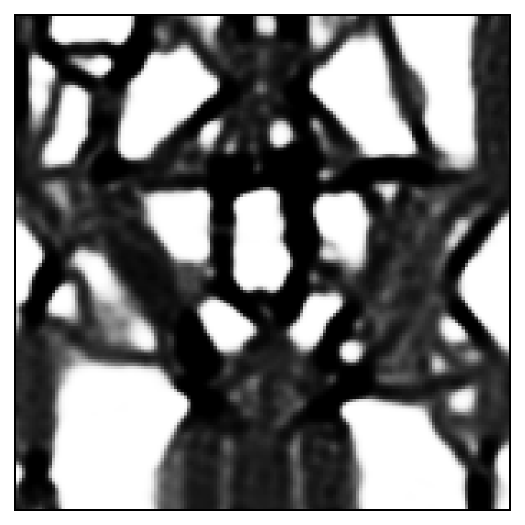}
    \caption{}
    \label{fig:designA}
  \end{subfigure}%
  \hfill
  \begin{subfigure}[b]{0.28\textwidth}
    \centering
    \includegraphics[width=\textwidth]{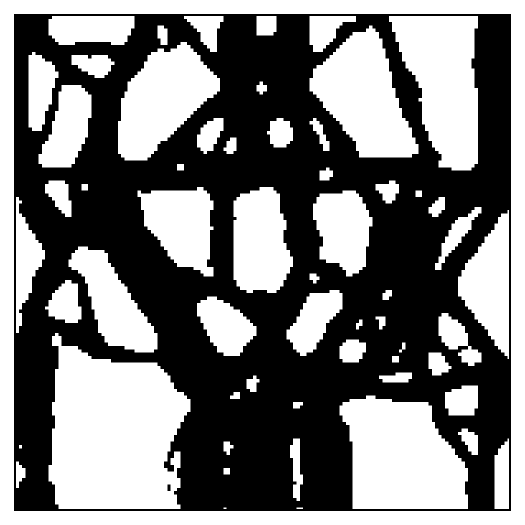}
    \caption{}
    \label{fig:designB}
  \end{subfigure}%
  \hfill
  \begin{subfigure}[b]{0.28\textwidth}
    \centering
    \includegraphics[width=\textwidth]{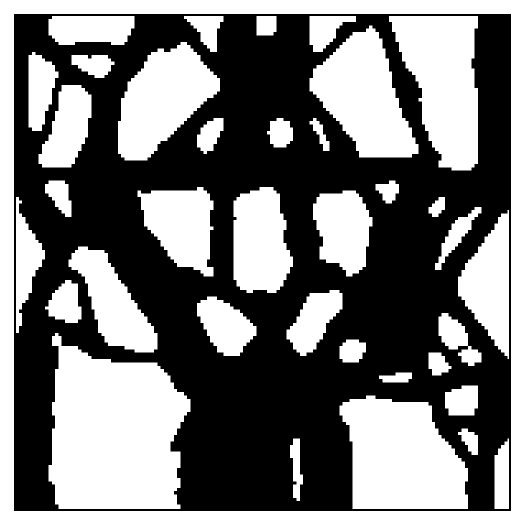}
    \caption{}
    \label{fig:designC}
  \end{subfigure}
  \caption{Representative 2D design evolution: (a) Optimized design obtained directly from the topology optimization process; (b) Binary design after applying a thresholding technique; (c) Final cleaned design after post-processing to remove tiny features. This design is used for subsequent finite element simulations and dataset generation.}
  \label{fig:three_designs}
\end{figure}
\subsection{GRU material model performance}

Overall, the GRU material models trained on the optimized-specimen dataset demonstrate strong predictive capability for history-dependent elastoplastic behavior. The performance of the GRU is quantified using the Normalized Root Mean Squared Error (NRMSE), defined as 

\begin{equation}
\mathrm{NRMSE}(\sigma_{ij}) =
\frac{\mathrm{RMSE}(\sigma_{ij})}{\mathrm{MAV}(\bar{\sigma}_{ij})}, 
\quad \{i,j\} = 1, \ldots, n_{\mathrm{dim}} ,
\end{equation}

where $\mathrm{RMSE}(\sigma_{ij})$ denotes the Root Mean Squared Error of stress component $\sigma_{ij}$ over all $n_t$ time steps, and $\mathrm{MAV}(\bar{\sigma}_{ij})$ represents the Mean Absolute Value of the corresponding ground-truth stress. Intuitively, the NRMSE measures the average prediction error relative to the typical magnitude of the stress response, thereby providing a scale-independent metric for model evaluation.

In the following sections, we first compare GRU models trained on datasets from the optimized specimen and from standard dogbone and notched specimens to assess the efficacy of the topology optimization framework. We then examine the influence of training set size on prediction accuracy and conduct a redundancy study to demonstrate that dataset diversity is a key factor for effective surrogate training. Finally, we investigate the generalization capability of the optimized specimen by generating training data using a Drucker–Prager plasticity model and evaluating GRU performance on the corresponding test set.

\subsubsection{Comparison with standard specimens: dogbone and notched geometries}

\begin{figure}[h!]
  \centering
  \begin{subfigure}[b]{0.32\textwidth}
    \centering
    \includegraphics[width=\textwidth]{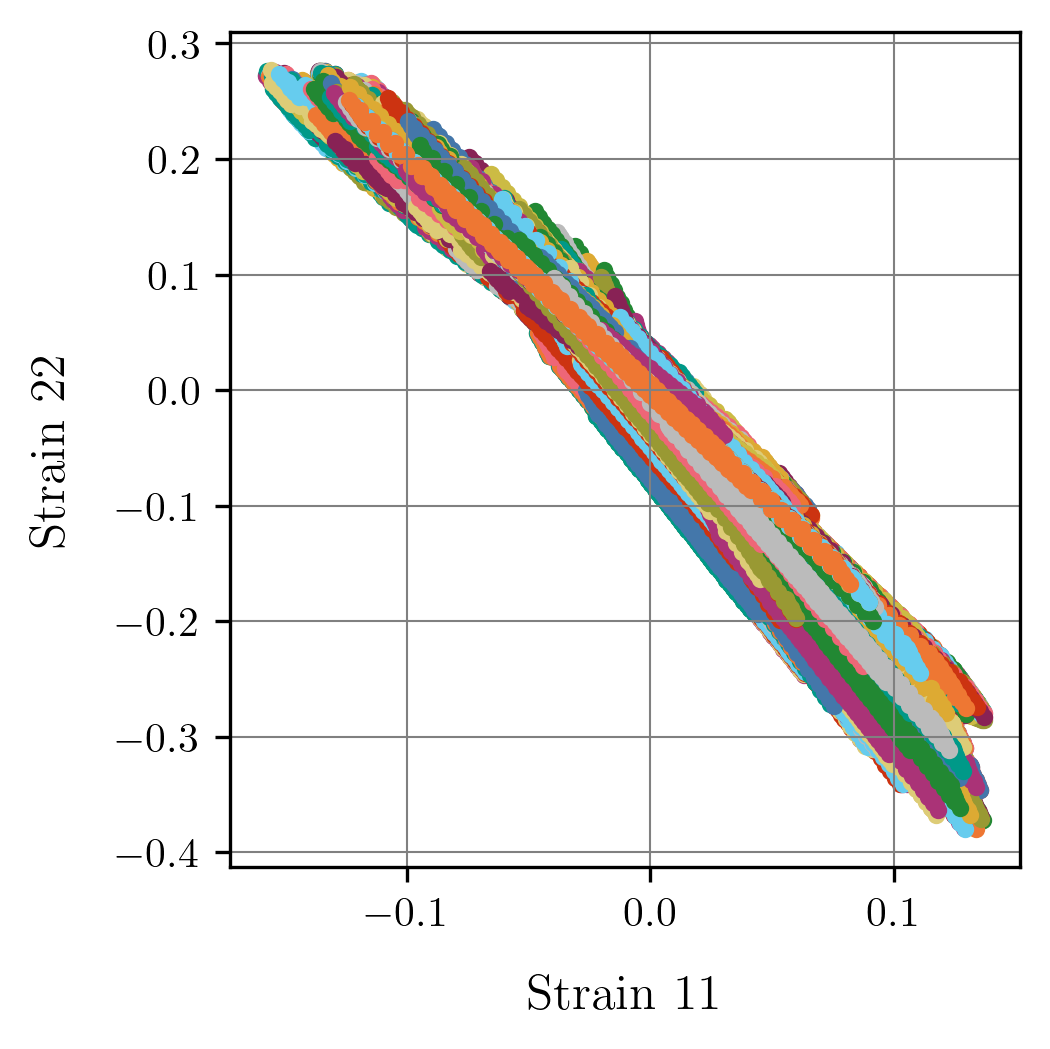}
    \caption{}
    \label{fig:11-22}
  \end{subfigure}%
  \hfill
  \begin{subfigure}[b]{0.32\textwidth}
    \centering
    \includegraphics[width=\textwidth]{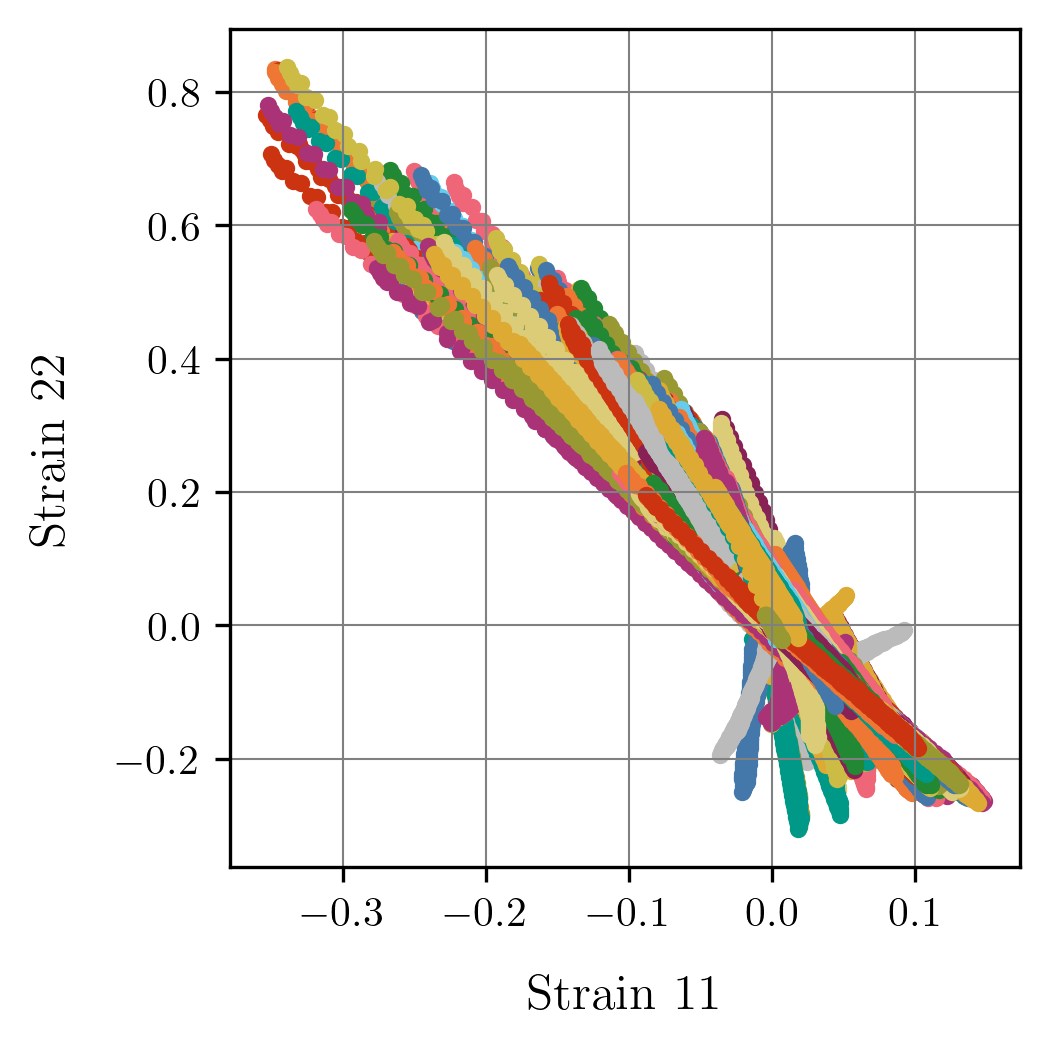}
    \caption{}
    \label{fig:11-12}
  \end{subfigure}%
  \hfill
  \begin{subfigure}[b]{0.32\textwidth}
    \centering
    \includegraphics[width=\textwidth]{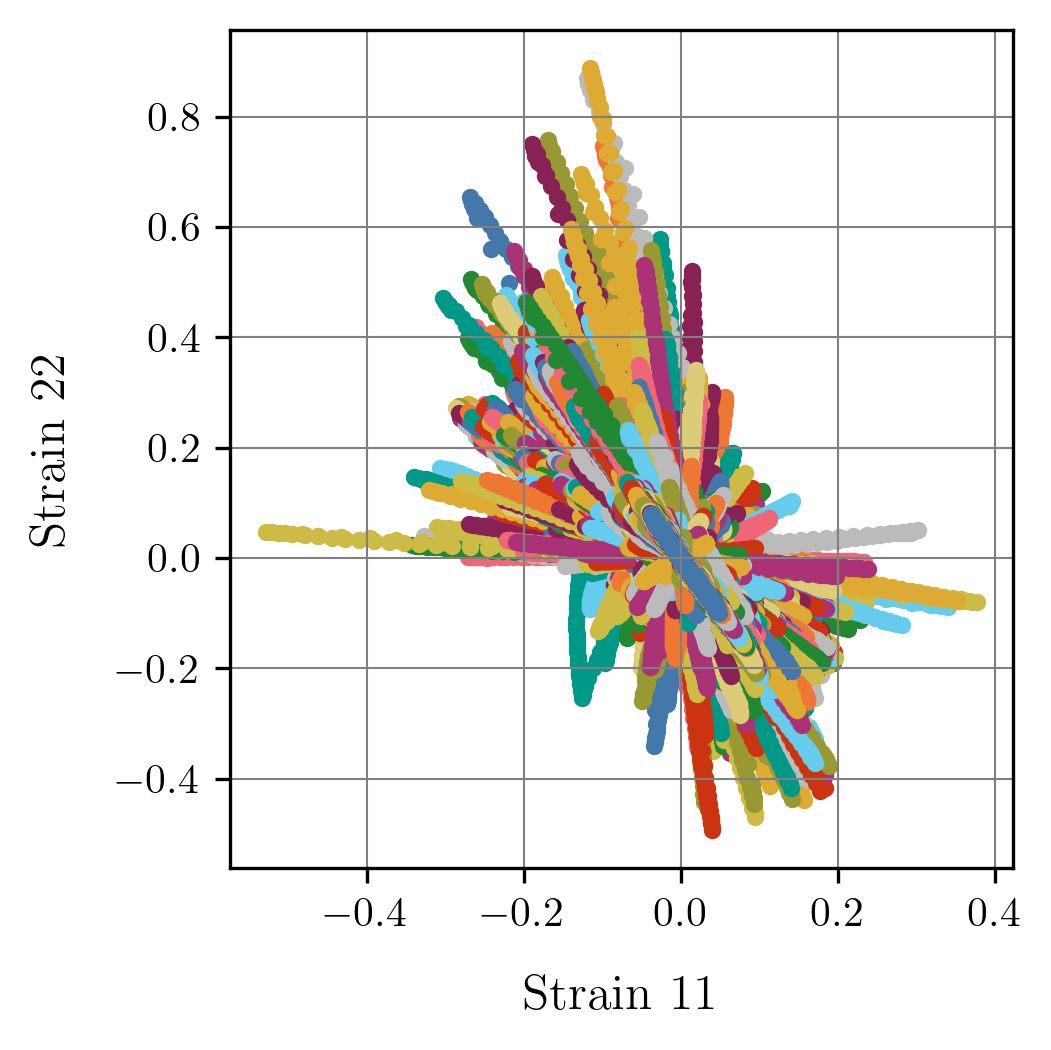}
    \caption{}
    \label{fig:22-12}
  \end{subfigure}
    \caption{Strain paths generated under two-cycle tension–compression loading for (a) the dogbone specimen, (b) the notched specimen, and (c) the optimized specimen. All paths are projected onto the $\varepsilon_{11}$–$\varepsilon_{22}$ space. Compared to the standard specimens, the optimized design produces more diverse strain paths.}

  \label{fig: specimen_strain_curves}
\end{figure}

Standard dogbone or notched specimens \cite{lou_general_2022} offer limited strain diversity for data-driven modeling (Fig. \ref{fig: specimen_strain_curves}), yielding high prediction errors (Tab. \ref{tab:prediction}). In contrast, our topology-optimized specimen maximizes strain path diversity under the same uniaxial cyclic loading (max displacement 10\% of length), covering a wider strain space. As a result, GRU models trained on this optimized dataset achieve significantly lower NRMSE across all stress components, demonstrating the necessity of heterogeneous geometry for surrogate training. (See Appendix \ref{app: loading_cases} for loading scenarios and Appendix \ref{app:random_designs} for distinct random design comparisons).

\begin{table}[h!]
\caption{Prediction performance for 2D stress components with different specimens.}
\label{tab:prediction}
\centering
\setlength{\tabcolsep}{0.5cm}
\renewcommand{\arraystretch}{1.3}
\begin{tabular}{cccc}
\toprule
\multirow{2}{*}{\textbf{Specimen}} 
& \multicolumn{3}{c}{\textbf{NRMSE (\%)}} \\
\cmidrule{2-4}
& \(\boldsymbol{\sigma_{11}}\) & \(\boldsymbol{\sigma_{22}}\) & \(\boldsymbol{\sigma_{12}}\) \\
\midrule
Optimized & \textbf{9.39} & \textbf{11.03} & \textbf{10.88} \\
Dogbone   & 87.38 & 166.36 & 55.52 \\
Notch     & 64.78 & 81.11  & 54.63 \\
\bottomrule
\end{tabular}
\end{table}

\subsubsection{Effect of training set size and redundancy on GRU performance}

\begin{figure}[h!]
    \centering
    \begin{subfigure}[t]{0.45\textwidth}
        \centering
        \includegraphics[width=\linewidth]{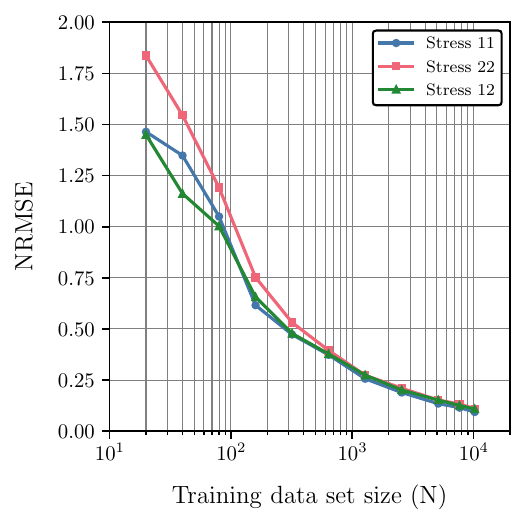}
        \caption{}
        \label{fig:nrmse_vs_training_size}
    \end{subfigure}
    \hfill
    \begin{subfigure}[t]{0.48\textwidth}
        \centering
        \includegraphics[width=\linewidth]{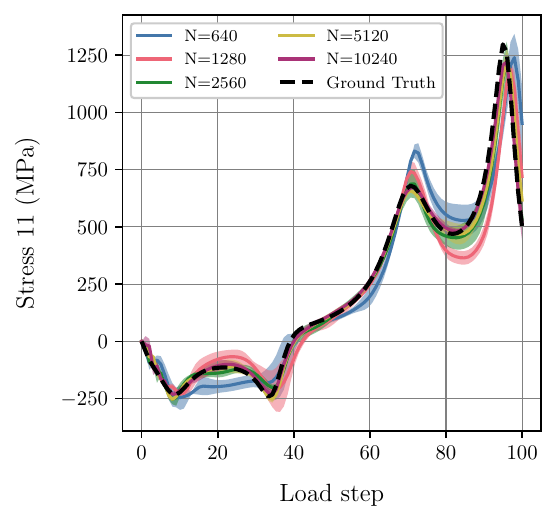}
        \caption{}
    \end{subfigure}
    \hfill
    \begin{subfigure}[t]{0.48\textwidth}
        \centering
        \includegraphics[width=\linewidth]{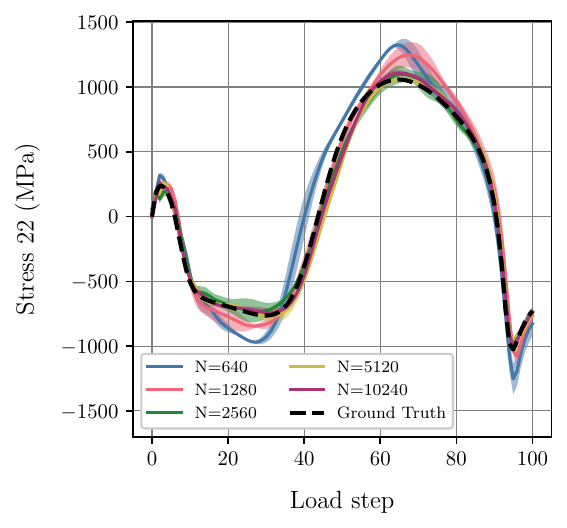}
        \caption{}
    \end{subfigure}
        \hfill
    \begin{subfigure}[t]{0.48\textwidth}
        \centering
        \includegraphics[width=\linewidth]{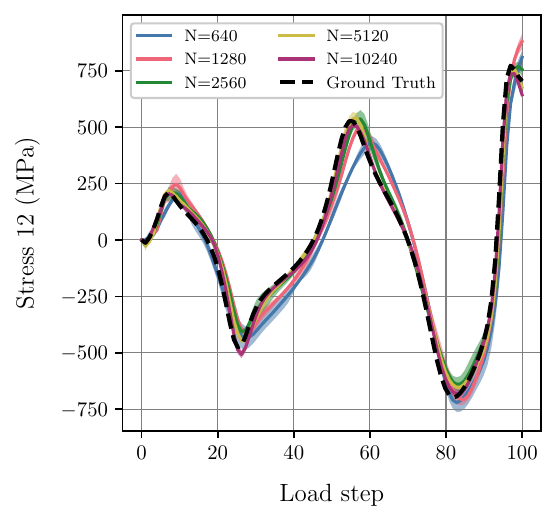}
        \caption{}
    \end{subfigure}
    \caption{(a) Prediction performance (NRMSE) with respect to the training data set size. (b–d) The mean and standard deviation (from 3 random model initializations) of predicted stress components on the random polynomial stress–strain paths. The displayed stress paths are selected based on having prediction errors closest to the mean error across the entire test set (reported in Tab. \ref{tab:prediction}), providing representative examples of model performance.}
    \label{fig:nrmse_and_stress_std}
\end{figure}

As mentioned in the Introduction, data-driven modeling with large neural networks requires large, diverse datasets, unlike traditional calibration. Examining the training set size and redundancy can help validate the optimized specimen's data quality. We sampled training sets of increasing size from the optimized specimen's data and trained independent GRU models. Fig. \ref{fig:nrmse_vs_training_size} confirms that accuracy improves with dataset size, while Fig. \ref{fig:nrmse_and_stress_std}(b-d) shows accurate, low-uncertainty predictions. We selected $10240$ paths for subsequent experiments.

\begin{figure}[ht]
  \centering
  \begin{subfigure}[b]{0.48\textwidth}
    \centering
    \includegraphics[width=\textwidth]{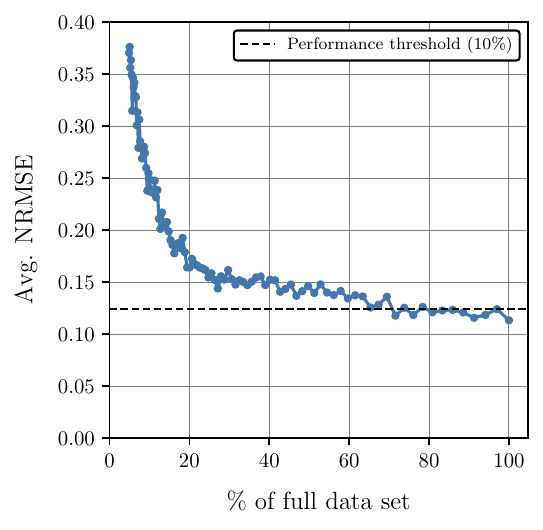}
    \caption{}
    \label{fig:top_loss}
  \end{subfigure}%
  \hfill
  \begin{subfigure}[b]{0.48\textwidth}
    \centering
    \includegraphics[width=\textwidth]{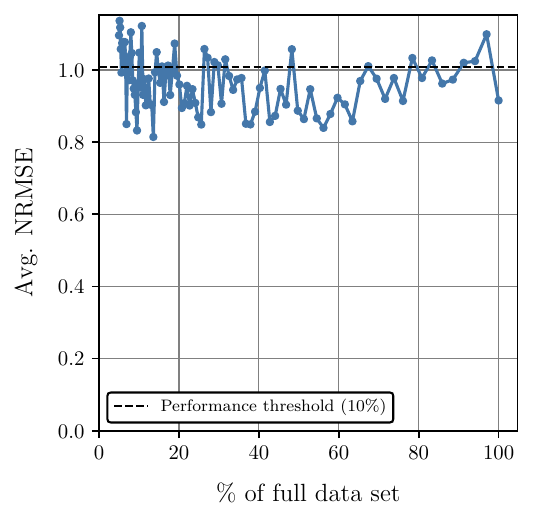}
    \caption{}
    \label{fig:iterations}
  \end{subfigure}%
  \caption{Dataset redundancy analysis. (a) Average prediction NRMSE (averaged over all stress components) of the GRU model, tested on an (unseen) randomly generated polynomial dataset.  The results show that training with only about $65\%$ of the optimized-specimen dataset achieves nearly the same accuracy as training with the full dataset, indicating relatively low redundancy. (b) The same study on the dogbone specimen. The curves are almost flat with huge prediction error, showing that the redundancy in the dogbone specimen is high, and there are not enough diverse stress–strain paths to train a GRU model.}
  \label{fig:pruning}
\end{figure}

To quantify redundancy, we performed a pruning study \cite{li_exploiting_2023}. A point is deemed redundant if removing it degrades test accuracy (NRMSE) by less than 10\%. Fig. \ref{fig:pruning} shows that while ~35\% of the optimized specimen's data is redundant, the dogbone specimen exhibits much higher redundancy, confirming the optimized design yields more informative data.

\subsubsection{Generalization of the optimized specimen to Drucker–Prager plasticity}

Real materials may deviate from the constitutive model assumed during optimization. To test generalizability, we simulated the specimen (designed using von Mises) using a Drucker–Prager model, keeping all other parameters constant (geometry, mesh, loading). We generated a new dataset of $10240$ paths and retrained the GRU model.

\begin{figure}[h!]
  \centering
  \begin{subfigure}[b]{0.32\textwidth}
    \centering
    \includegraphics[width=\textwidth]{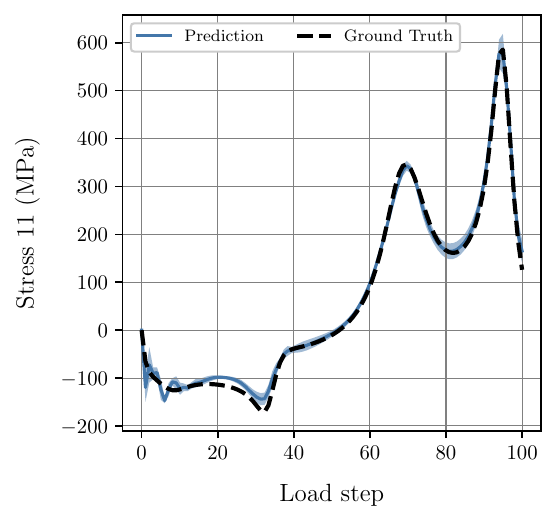}
    \caption{}
  \end{subfigure}%
  \hfill
  \begin{subfigure}[b]{0.32\textwidth}
    \centering
    \includegraphics[width=\textwidth]{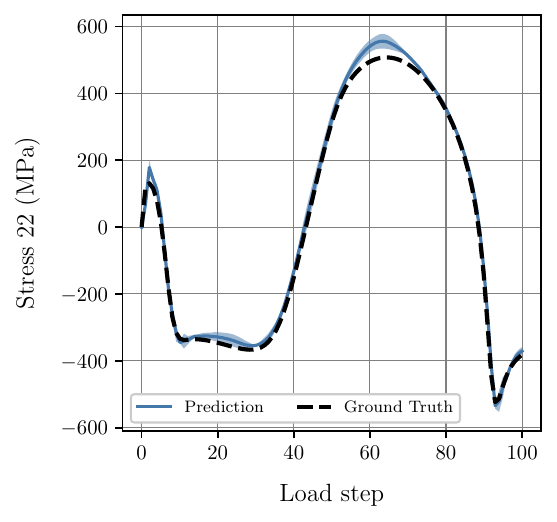}
    \caption{}
  \end{subfigure}%
  \hfill
  \begin{subfigure}[b]{0.32\textwidth}
    \centering
    \includegraphics[width=\textwidth]{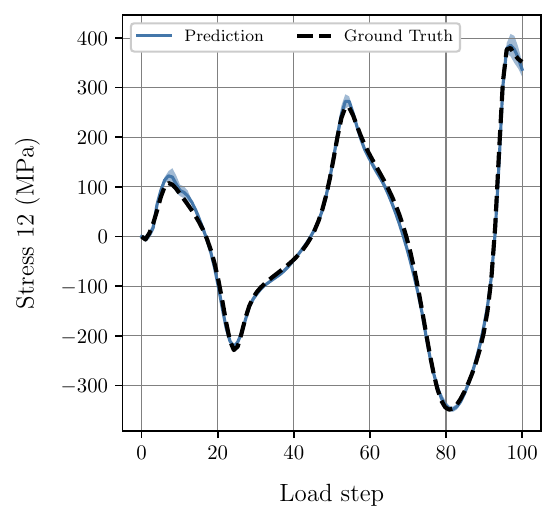}
    \caption{}
  \end{subfigure}
  \caption{The mean and standard deviation (from 3 random model initializations) of predicted stress components on the random polynomial stress–strain paths. The displayed stress paths are the same as those in the Von-Mises case.}
  \label{fig:dp_prediction}
\end{figure}

As shown in Tab. \ref{tab:prediction_DP} (NRMSE) and Fig. \ref{fig:dp_prediction} (predictions), the GRU accurately captures Drucker–Prager behavior. The small performance drop compared to von Mises is due to reduced strain path diversity from pressure sensitivity. Nevertheless, this proves the optimized specimen generates informative data even when the underlying material model changes, supporting its robustness and future applicability to real experimental data.

\begin{table}[h!]
\caption{Prediction performance for 2D stress components with Drucker--Prager model.}
\label{tab:prediction_DP}
\centering
\setlength{\tabcolsep}{0.5cm}
\renewcommand{\arraystretch}{1.3}
\begin{tabular}{lccc}
\toprule
\multirow{2}{*}{\textbf{Material model}}  & \multicolumn{3}{c}{\textbf{NRMSE (\%)}} \\
\cmidrule(lr){2-4}
& \(\boldsymbol{\sigma_{11}}\) & \(\boldsymbol{\sigma_{22}}\) & \(\boldsymbol{\sigma_{12}}\) \\
\midrule
Drucker--Prager & 10.56 & 13.05 & 10.80 \\
\bottomrule
\end{tabular}
\end{table}

\section{Discussion}

This study aims to identify an optimized specimen design for training neural network surrogates, but we acknowledge that the obtained design is not guaranteed to be globally optimal. The optimized specimen geometry could vary if certain parameters within the design framework were altered. The central idea here is to maximize strain-state diversity as much as possible. While it is likely that GRU performance could be further improved by tuning hyperparameters in both the topology optimization framework and the material model updating framework, such refinements are outside the scope of the present work. Instead, this study focuses on demonstrating the overall workflow and establishing the feasibility of integrating topology optimization with data-driven surrogate modeling.

\begin{figure}[h!]
\centering
\includegraphics[width=0.99\textwidth]{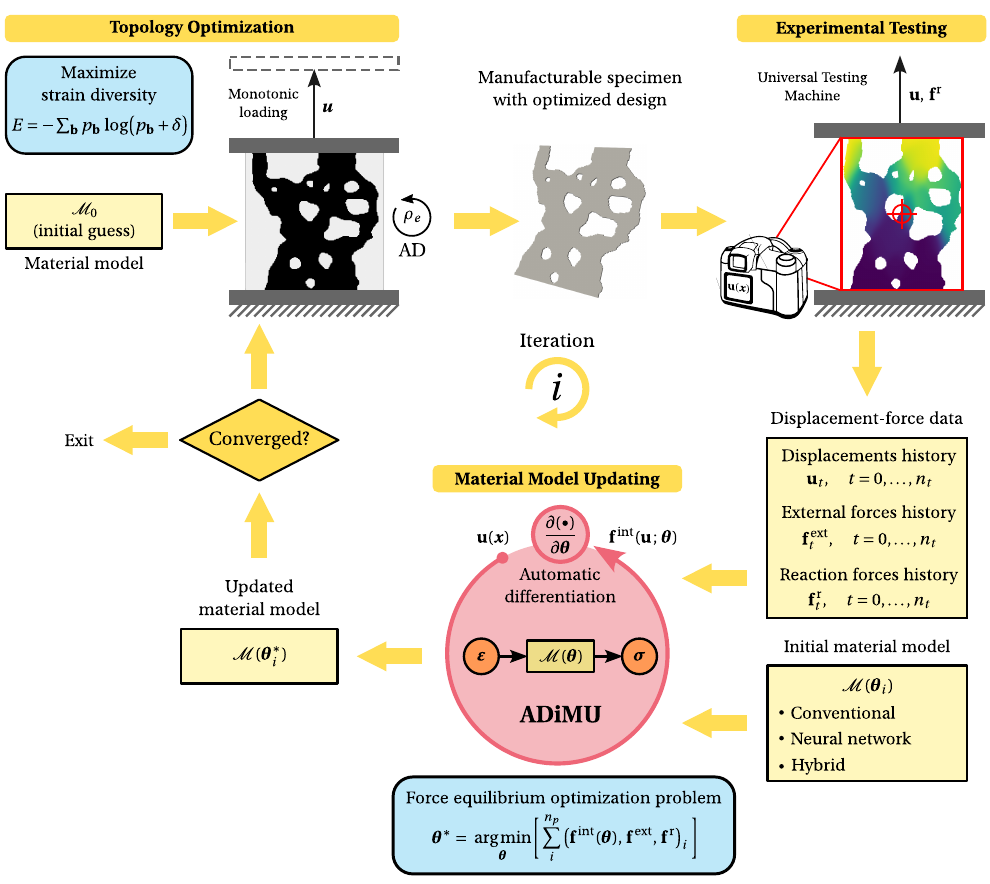}
\caption{Proposed experimental-data-driven topology optimization workflow for material model updating. The procedure consists of three main stages. First, automatic differentiation–enhanced elastoplastic topology optimization is employed to maximize strain-state diversity under simple monotonic loading, resulting in a manufacturable, optimized specimen geometry. Second, this optimized specimen is fabricated and experimentally tested under controlled loading conditions, capturing the full-field displacement and reaction-force data. Then, experimental data are integrated into the automatic-differentiation-based material model updating framework (ADiMU) to either calibrate conventional constitutive models or train neural network surrogate models. Finally, the newly calibrated or trained material model can subsequently be used within the topology optimization framework, iteratively improving specimen designs based on experimental data, ultimately enabling accurate constitutive modeling from a single, carefully designed mechanical test.}
\label{fig:futurework}
\end{figure}

Although this study is fully computational, several of the optimized specimens are feasible for manufacturing and experimental validation. In future work, we plan to fabricate these designs and experimentally measure full-field displacement responses under controlled loading conditions using Digital Image Correlation (DIC). Unlike the present study, which relies on local stress–strain paths, such experiments will provide full-field displacement data with global force data.

By leveraging the global material model updating part in the ADiMU framework, the experimental data obtained from these specimens can be used not only to calibrate traditional constitutive models but also to train GRU-based recurrent neural network models. Subsequently, these experimentally informed machine learning models can be directly incorporated back into the topology optimization process, thereby replacing the initially assumed constitutive laws with data-driven models derived from actual global experimental responses. This establishes a self-consistent loop as shown in Fig. \ref{fig:futurework}: 
\begin{enumerate}
    \item The topology optimization framework creates specimens generating diverse strain states;
    \item Experiments provide global displacement and force measurements;
    \item The ADiMU framework calibrates or trains models using these global data;
    \item The resulting models are reincorporated into further optimization.
\end{enumerate}
Through iterative refinement of specimen designs guided by experimental data, we envision realizing the ultimate objective: accurately characterizing material constitutive behavior from a single, well-designed uniaxial test.

\section{Conclusion}

In this study, we propose an automatic differentiation enhanced elastoplastic topology optimization approach for designing mechanical specimens capable of generating diverse stress–strain datasets under simple uniaxial loading conditions. We formulate an entropy-based objective function to quantify and maximize the spread of local strain states. Both 2D and 3D optimized specimens are created, and their effectiveness is evaluated by training and testing GRU-based recurrent neural network models using specimen-derived datasets. The results show that GRU models trained on data from optimized specimens significantly outperform those trained on datasets from random or standard dogbone/notched geometries, as validated on randomly generated polynomial datasets. Additionally, a pruning analysis revealed substantially lower data redundancy within optimized specimen datasets compared to standard dogbone datasets, highlighting the critical role of strain-path diversity in effectively training fully data-driven surrogate models. We also evaluate the robustness and generalizability of the optimization framework by testing the optimized specimen with a Drucker-Prager model. To our knowledge, this work represents the first integration of elastoplastic heterogeneous specimen design via topology optimization with recurrent neural network surrogate material modeling. 

\section{Acknowledgments}

The authors would like to acknowledge that this effort was undertaken in part with the support from the Department of the Navy, Office of Naval Research, award number N00014-23-1-2688. The corresponding author acknowledges the use of a large language model (Gemini 3) to edit the text of this manuscript after a complete draft was written.

\newpage
\appendix

\section{Supplementary Results: 2D Plane-Stress Case}
\subsection{Effect of filter size on optimized specimen geometry and GRU performance}
\label{app:filter}

\begin{figure}[h!]
  \centering
  \begin{subfigure}[b]{0.24\textwidth}
    \centering
    \includegraphics[width=\textwidth]{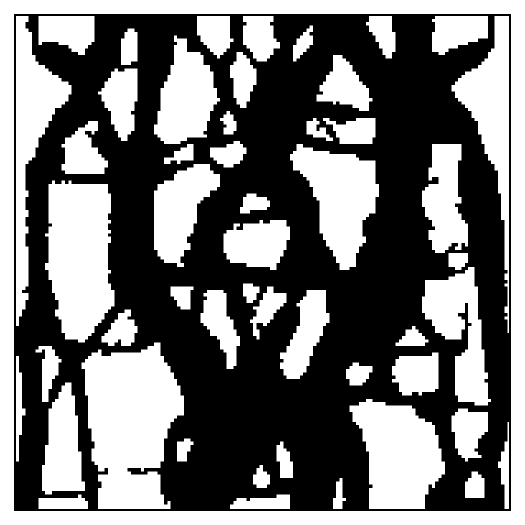}
    \caption{}
  \end{subfigure}
  \hfill
  \begin{subfigure}[b]{0.24\textwidth}
    \centering
    \includegraphics[width=\textwidth]{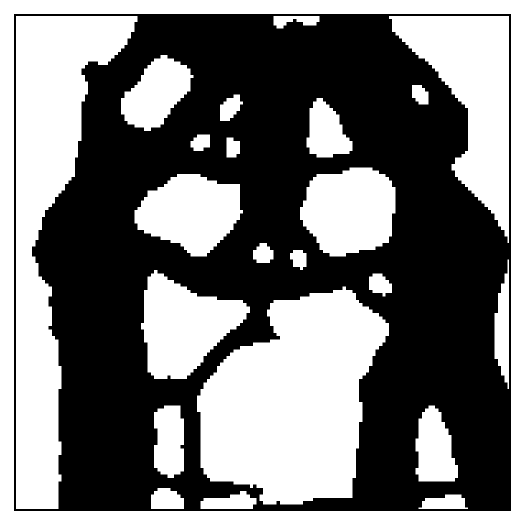}
    \caption{}
  \end{subfigure}
  \hfill
  \begin{subfigure}[b]{0.24\textwidth}
    \centering
    \includegraphics[width=\textwidth]{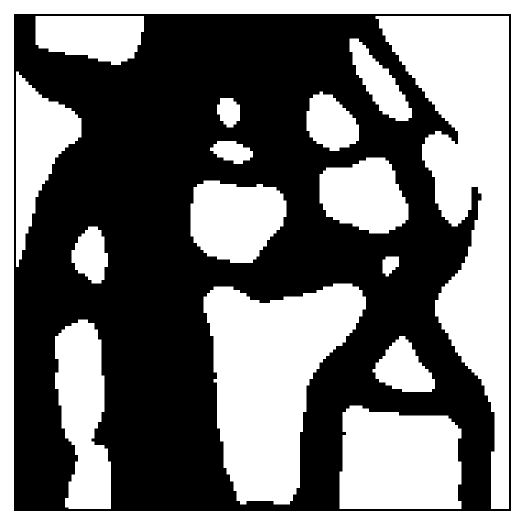}
    \caption{}
  \end{subfigure}
  \hfill
  \begin{subfigure}[b]{0.24\textwidth}
    \centering
    \includegraphics[width=\textwidth]{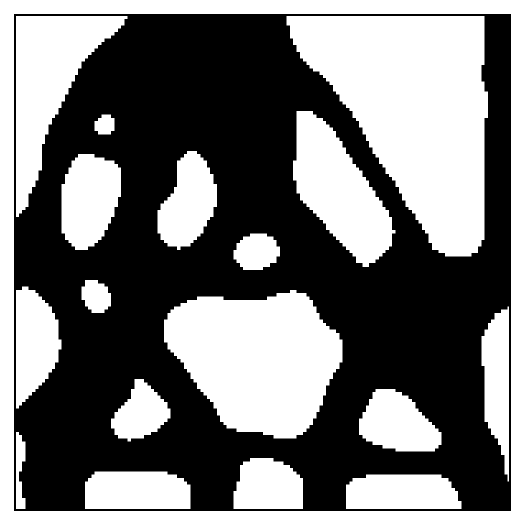}
    \caption{}
  \end{subfigure}
  \caption{Optimized 2D designs obtained using different filter radii ($r = 
 2,6,8,10$). As the filter size increases, the resulting geometry becomes smoother and coarser, reducing fine structural features. These variations in structural resolution influence the diversity of local strain paths and, consequently, the performance of GRU models trained on the resulting datasets. Design with $r=4$ is in the main paper.}
  \label{fig:filter_2d}
\end{figure}

The influence of filter size on the optimized specimen geometry is being evaluated for a fixed mesh discretization of $150 * 150$ elements. The filter size, ranging from 2 to 10 elements, controls the minimum feature size in the optimized specimen, thereby enforcing manufacturing constraints. Larger filter sizes facilitate compliance with fabrication requirements but may reduce the performance of GRU models trained on the resulting datasets. This reduction is attributed to the suppression of fine geometric features, which can generate unique stress-strain paths that enhance dataset diversity. In this study, five distinct filter sizes (2, 4, 6, 8, and 10) are being applied to generate optimized specimens. Corresponding datasets are produced for each specimen to train GRU models. The optimized geometries and their prediction performance NRMSE (Normalized Root Mean Squared Error) for 2D plane stress simulations are presented in Fig. \ref{fig:filter_2d} and Tab. \ref{tab:filter_radius_nrmse_2d}.

\begin{table}[h!]
\caption{Prediction performance (NRMSE) of GRU models trained on strain–stress data generated from 2D optimized specimens with different filter radius elements.}
\label{tab:filter_radius_nrmse_2d}
\centering
\setlength{\tabcolsep}{0.30cm}
\renewcommand{\arraystretch}{1.5}
\newcolumntype{C}[1]{>{\centering\arraybackslash}p{#1}}
\begin{tabular}{C{3cm} C{2cm} C{2cm} C{2cm}}

\toprule
\multirow{2}{*}{\parbox[c]{1.5cm}{\centering \bfseries \shortstack{Filter \\[0.3em] Radius}}}
  & \multicolumn{3}{c}{\centering \bfseries NRMSE (\%)} \\
\cmidrule(l){2-4}
  & \bfseries $\sigma_{11}$ & \bfseries $\sigma_{22}$ & \bfseries $\sigma_{12}$ \\
\midrule
Filter\ 2 & 9.78 & 11.52 & 11.11 \\ \midrule
Filter\ 4 & 9.39 & 11.03 & 10.88 \\ \midrule
Filter\ 6 & 10.35 & 12.18 & 12.06 \\ \midrule
Filter\ 8 & 11.36 & 13.24 & 13.18 \\ \midrule
Filter\ 10 & 11.31 & 12.94 & 13.23 \\ 
\bottomrule
\end{tabular}
\end{table}

\subsection{Effect of training loading conditions on GRU performance}
\label{app: loading_cases}
This section evaluates the performance of GRU models being trained on datasets generated from various loading scenarios applied to the optimized specimen. The loading conditions under investigation include monotonic tension (1), tension with loading and unloading (2), one-cycle tension-compression (3), and two-cycle tension-compression (4). Additionally, four combined scenarios are being considered, each pairing a tension-dominated case with its compression-dominated counterpart in equal proportions: monotonic tension with monotonic compression (5), tension with loading/unloading with compression with loading/unloading (6), one-cycle tension-compression with one-cycle compression-tension (7), and two-cycle tension-compression with two-cycle compression-tension (8). These combined datasets are constructed such that half of the data is from tension-dominated loading and the other half from compression-dominated loading, resulting in eight distinct loading cases, each comprising $10240$ training data paths. For each loading case, GRU models with three random initializations are trained and tested, and the average prediction performance is shown in Table \ref{tab:loading}. It demonstrates that the eighth loading case, which integrates both two-cycle tension-compression and two-cycle compression-tension scenarios, yields the highest performance for GRU training. However, this case requires two separate tests on the same specimen to generate the complete dataset, which conflicts with the objective of utilizing a single test. Consequently, the two-cycle tension-compression loading is being adopted for this study.

\begin{table}[h!]
\caption{Prediction performance (NRMSE) of stress components on the random polynomial test set for GRU models trained on datasets from eight loading cases on the optimized design. Loading 1: Monotonic Tension; Loading 2: Tension with Loading/Unloading; Loading 3: One-Cycle Tension-Compression; Loading 4: Two-Cycle Tension-Compression; Loading 5: Monotonic Tension + Compression; Loading 6: Tension + Compression with Loading/Unloading; Loading 7: One-Cycle Tension-Compression + Compression-Tension; Loading 8: Two-Cycle Tension-Compression + Compression-Tension. }
\label{tab:loading}
\centering
\small
\setlength{\tabcolsep}{0.30cm}
\newcolumntype{C}[1]{>{\centering\arraybackslash}p{#1}}
\renewcommand{\arraystretch}{1.5}
\begin{tabular}{C{3cm} C{2cm} C{2cm} C{2cm}}
\toprule
\multirow{2}{*}{\parbox[c]{1.5cm}{\centering \bfseries \shortstack{Loading \\[0.3em] Cases}}}
  & \multicolumn{3}{c}{\centering \bfseries NRMSE (\%)} \\ 
\cmidrule(l){2-4}
  & \bfseries $\sigma_{11}$ & \bfseries $\sigma_{22}$ & \bfseries $\sigma_{12}$ \\ 
\midrule
Loading\ 1 & 169.44 & 180.31 & 176.70 \\ \midrule
Loading\ 2 & 102.05 & 103.47 & 104.29 \\ \midrule
Loading\ 3 &  29.41 &  34.18 &  32.27 \\ \midrule
Loading\ 4 &  \cellcolor{lightgray} 9.49 & \cellcolor{lightgray} 10.50 & \cellcolor{lightgray} 10.64 \\ \midrule
Loading\ 5 & 150.35 & 144.57 & 158.42 \\ \midrule
Loading\ 6 &  93.48 &  90.53 &  96.15 \\ \midrule
Loading\ 7 &  15.52 &  18.12 &  18.09 \\ \midrule
Loading\ 8 &   8.91 &   9.94 &   9.89 \\
\bottomrule
\end{tabular}
\end{table}

\subsection{GRU performance trained on randomly generated specimens}
\label{app:random_designs}

\begin{figure}[h!]
  \centering
  \begin{subfigure}[b]{0.19\textwidth}
    \centering
    \includegraphics[width=\textwidth]{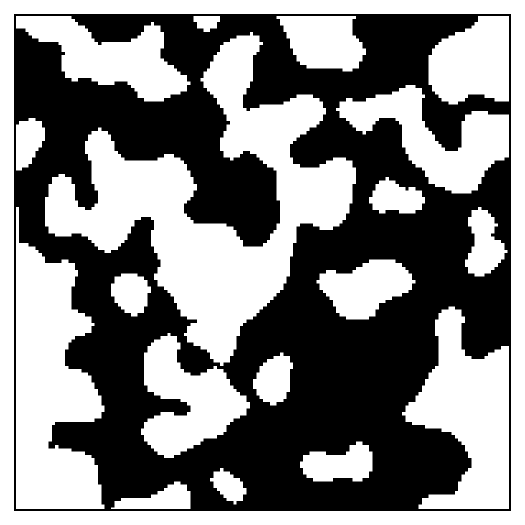}
    \caption{}
  \end{subfigure}
  \hfill
  \begin{subfigure}[b]{0.19\textwidth}
    \centering
    \includegraphics[width=\textwidth]{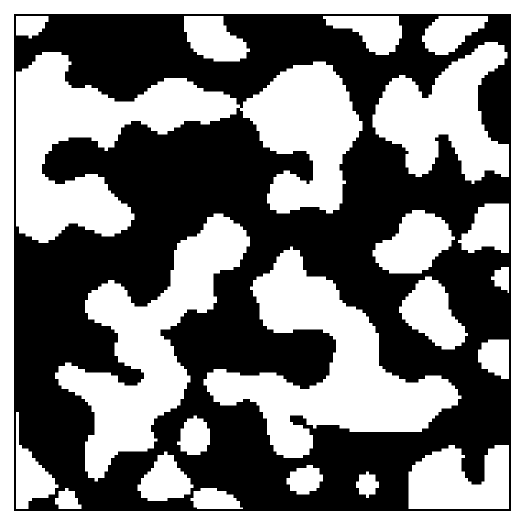}
    \caption{}
  \end{subfigure}
  \hfill
  \begin{subfigure}[b]{0.19\textwidth}
    \centering
    \includegraphics[width=\textwidth]{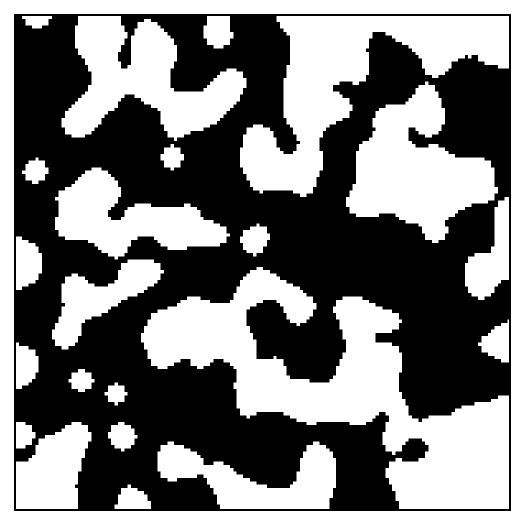}
    \caption{}
  \end{subfigure}
    \hfill
  \begin{subfigure}[b]{0.19\textwidth}
    \centering
    \includegraphics[width=\textwidth]{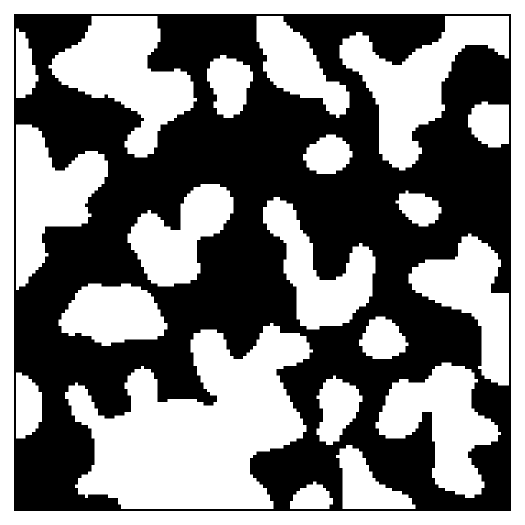}
    \caption{}
  \end{subfigure}
    \hfill
  \begin{subfigure}[b]{0.19\textwidth}
    \centering
    \includegraphics[width=\textwidth]{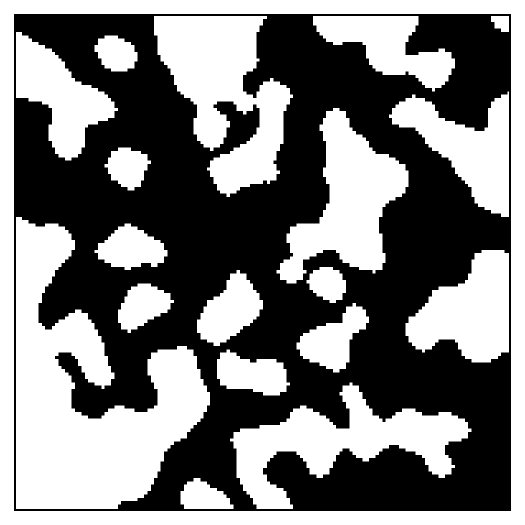}
    \caption{}
  \end{subfigure}
  \caption{Randomly generated designs.}
  \label{fig:random_designs}
\end{figure}

In this section, we evaluate the effectiveness of the optimized specimen by comparing it with five randomly generated designs. All the specimens are subjected to the same loading conditions to generate stress-strain data for training GRU models. The random designs are constructed using the following procedure: first, a $150 \times 150$ grid of random noise is generated and smoothed using a Gaussian filter to introduce spatial correlation. Second, the smoothed field is thresholded to produce a binary design with a volume fraction comparable to that of the topology-optimized design. Third, a morphological closing operation is applied to fill small gaps and enhance connectivity. Finally, only the largest connected solid component is retained to form the final design, as illustrated in Fig. \ref{fig:random_designs}.

The parameters used to generate these random designs are carefully tuned to ensure that the resulting designs have similar structural features to those of the topology-optimized specimen, allowing for fair comparison. However, as shown in Tab. \ref{tab:random_designs}, the GRU models trained on datasets derived from the random designs perform significantly worse than those trained on the topology-optimized specimen. Moreover, the uncertainty across the five random designs is substantial due to the uncontrolled geometries.

\begin{table}[h!]
\caption{Prediction performance (NRMSE) of GRU models trained on strain–stress data generated from randomly designed specimens.}
\label{tab:random_designs}
\centering
\small
\setlength{\tabcolsep}{0.30cm}
\newcolumntype{C}[1]{>{\centering\arraybackslash}p{#1}}
\renewcommand{\arraystretch}{1.5}
\begin{tabular}{C{3cm} C{2cm} C{2cm} C{2cm}}
\toprule
\multirow{2}{*}{\parbox[c]{1.5cm}{\centering \bfseries \shortstack{Random \\[0.3em] Designs}}}
  & \multicolumn{3}{c}{\centering \bfseries NRMSE (\%)} \\ 
\cmidrule(l){2-4}
  & \bfseries $\sigma_{11}$ & \bfseries $\sigma_{22}$ & \bfseries $\sigma_{12}$ \\ 
\midrule
Random\ Design\ 1 & 22.10 & 28.93 & 29.47 \\ \midrule
Random\ Design\ 2 & 42.16 & 54.03 & 51.43 \\ \midrule
Random\ Design\ 3 & 22.26 & 27.17 & 25.86 \\ \midrule
Random\ Design\ 4 & 17.40 & 21.14 & 21.69 \\ \midrule
Random\ Design\ 5 & 25.45 & 24.95 & 25.14 \\ \midrule
\textbf{Mean} & \textbf{25.87} & \textbf{31.64} & \textbf{30.32} \\
\bottomrule
\end{tabular}
\end{table}

\subsection{GRU performance trained on standard specimens}
\label{app:standard_specimen}

In this section, the standard specimens, including a dogbone and a notched specimen, are used to generate stress-strain paths under the same loading conditions as those applied in the optimized design. The geometric configuration follows the setup described in Ref. \cite{lou_general_2022}.  To ensure the extracted data are representative of the material response, only the central regions of the specimens are considered. From this region, a total of 6,000 stress-strain paths for the dogbone specimen and 2,000 for the notched specimen are collected and used to train the GRU model. As summarized in Tab. \ref{tab:dogbone}, the predictive performance based on the standard specimens is significantly worse than that obtained from the optimized specimen, even with fewer training paths, highlighting the effectiveness of our design strategy in generating diverse and informative training data.  \\

\begin{table}[h!]
\caption{Comparison of prediction performance (NRMSE) of GRU models trained on strain–stress data generated from a dogbone specimen, a notched specimen, and the topology-optimized specimen.}

\label{tab:dogbone}
\centering
\setlength{\tabcolsep}{0.30cm}
\newcolumntype{C}[1]{>{\centering\arraybackslash}p{#1}}
\renewcommand{\arraystretch}{1.5}
\begin{tabular}{C{3cm} C{3cm} C{2cm} C{2cm} C{2cm}}
\toprule
\multirow{2}{*}{\parbox[c]{1.5cm}{\centering \bfseries Specimens}}
  & \multirow{2}{*}{\parbox[c]{1.5cm}{\centering \bfseries Number \\[0.3em] of Paths}}
  & \multicolumn{3}{c}{\bfseries NRMSE (\%)} \\
\cmidrule(l){3-5}
  & 
  & \bfseries $\sigma_{11}$ & \bfseries $\sigma_{22}$ & \bfseries $\sigma_{12}$ \\
\midrule
dogbone   & 6000 & 87.38 & 166.36 & 55.52 \\ \midrule
notch     & 2000 & 64.78 & 81.11  & 54.63 \\ \midrule
optimized & 1280 & 25.66 & 27.36  & 27.44 \\
\bottomrule
\end{tabular}
\end{table}

\subsection{Mesh Dependency Study}
\label{app:mesh_dependent}

\begin{figure}[h!]
  \centering
  \begin{subfigure}[b]{0.32\textwidth}
    \centering
    \includegraphics[width=\textwidth]{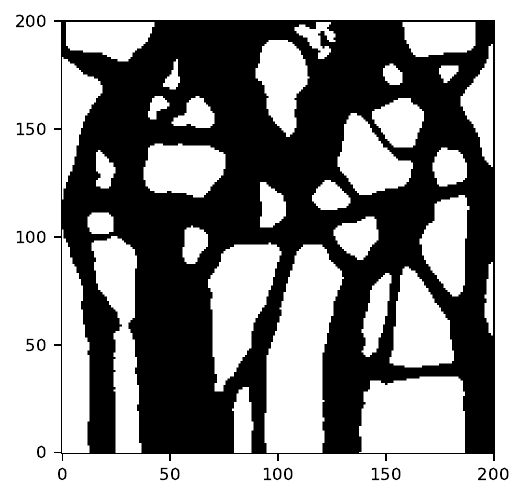}
    \caption{}
    \label{fig:200}
  \end{subfigure}%
  \hfill
  \begin{subfigure}[b]{0.32\textwidth}
    \centering
    \includegraphics[width=\textwidth]{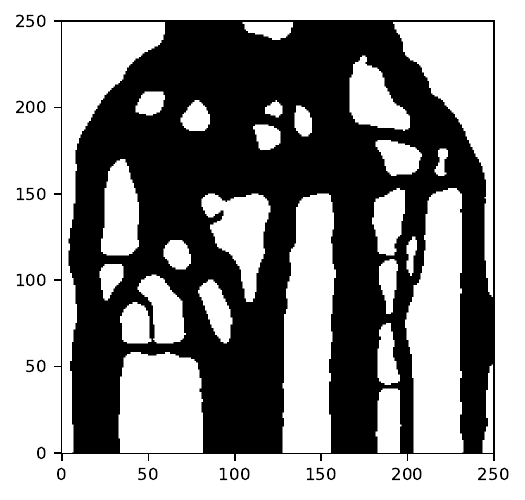}
    \caption{}
    \label{fig:250}
  \end{subfigure}%
  \hfill
  \begin{subfigure}[b]{0.32\textwidth}
    \centering
    \includegraphics[width=\textwidth]{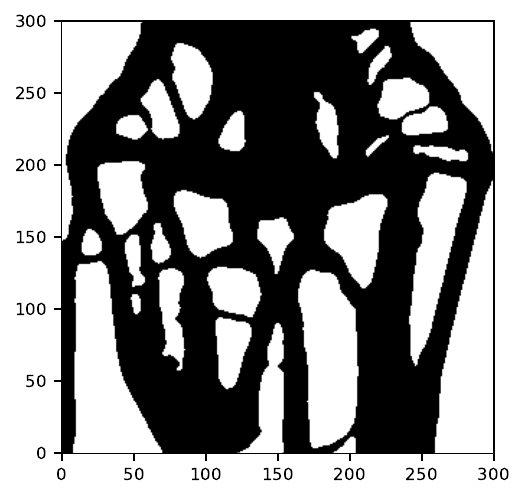}
    \caption{}
    \label{fig:300}
  \end{subfigure}
  \caption{Optimized designs for different mesh sizes. (a) In a $200*200$ mesh. (b) In a $250*250$ mesh. (c) In a $300*300$ mesh.}
  \label{fig:meshsize}
\end{figure}

This section studies the mesh dependency of the topology optimization framework. The entropy-based objective function incorporates local strain states from each element to construct a measure of strain space coverage. Consequently, finer meshes, with more elements, increase the likelihood of occupying predefined cells in the strain space, rendering the objective function mesh-dependent. To investigate this effect, the framework generates optimized specimen geometries for various mesh sizes, including $200*200$, $250*250$, and $300*300$. The filter size adjusts proportionally to the mesh refinement to maintain a consistent minimum feature size across all designs. Fig. \ref{fig:meshsize} illustrates the optimized geometries for different mesh sizes. In the following training process, the same number of stress-strain paths (10240) is randomly selected from these specimen datasets. Tab. \ref{tab:mesh} summarizes the prediction performance of GRU models on the same random polynomial testing set.

\begin{table}[h!]
\caption{Prediction performance (NRMSE) of GRU models trained on stress–strain data generated from topology-optimized designs with different meshes.}
\label{tab:mesh}
\centering
\setlength{\tabcolsep}{0.30cm}
\newcolumntype{C}[1]{>{\centering\arraybackslash}p{#1}}
\renewcommand{\arraystretch}{1.5}
\begin{tabular}{C{3cm} C{2cm} C{2cm} C{2cm}}
\toprule
\multirow{2}{*}{\parbox[c]{1.5cm}{\centering \bfseries Mesh \\ [0.3em] Resolutions}}
& \multicolumn{3}{c}{\bfseries NRMSE (\%)} \\
\cmidrule(l){2-4}
  & \bfseries $\sigma_{11}$ & \bfseries $\sigma_{22}$ & \bfseries $\sigma_{12}$ \\
\midrule
200*200     & 9.63  & 10.68  & 11.04 \\ \midrule
250*250     & 12.34 & 14.03  & 12.75 \\ \midrule
300*300     & 10.94 & 12.24  & 12.00 \\
\bottomrule
\end{tabular}
\end{table}

\section{Supplementary Results: 3D Case}
\label{app:3D_results}
Following the same approach as in the 2D plane stress simulations, the 3D design domain is constructed with identical in-plane dimensions ($L=10 \ \mathrm{mm}$) and mesh resolution ($150 \times 150$). The domain includes a single layer of elements in the out-of-plane direction, with a thickness of $L_z = 10/150 \ \mathrm{mm}$. A uniaxial displacement loading is applied on the top surface, while the bottom surface is fixed in the loading direction. Considering the thin structural thickness and the need for computational efficiency, only the diversity of the in-plane strain components is maximized in the topology optimization framework. Accordingly, the objective function is defined in the same manner as in the 2D plane stress case.

\begin{figure}[h!]
  \centering
  \begin{subfigure}[b]{0.32\textwidth}
    \centering
    \includegraphics[width=\textwidth]{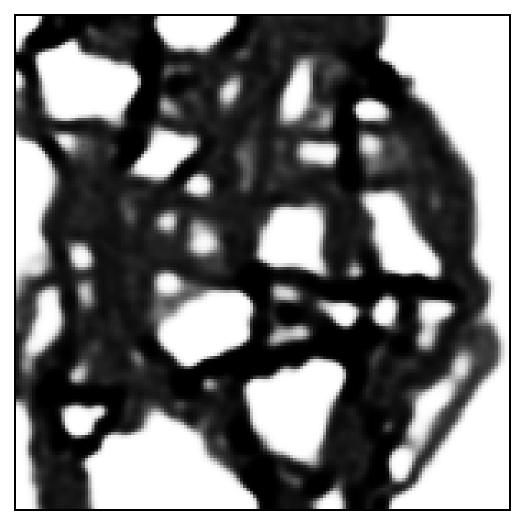}
    \caption{}
    \label{fig:designA_3D}
  \end{subfigure}%
  \hfill
  \begin{subfigure}[b]{0.32\textwidth}
    \centering
    \includegraphics[width=\textwidth]{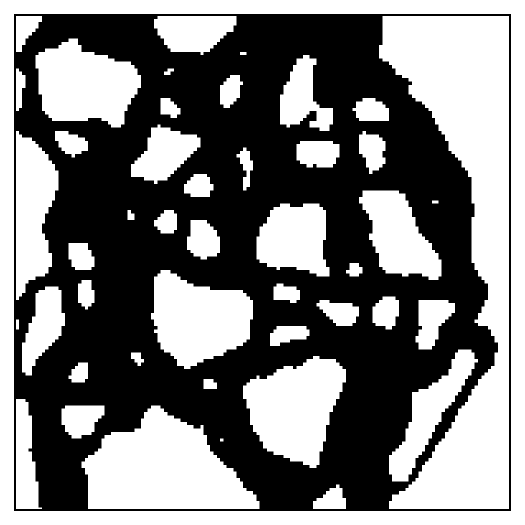}
    \caption{}
    \label{fig:designB_3D}
  \end{subfigure}%
  \hfill
  \begin{subfigure}[b]{0.32\textwidth}
    \centering
    \includegraphics[width=\textwidth]{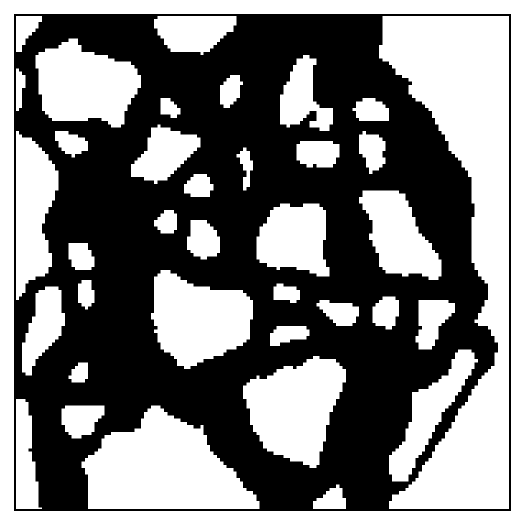}
    \caption{}
    \label{fig:designC_3D}
  \end{subfigure}
  \caption{Representative 3D design evolution. (a) Optimized design obtained directly from the topology optimization process. (b) Binary design after applying a thresholding technique. (c) Final cleaned design after post-processing to remove tiny features. This design is used for subsequent finite-element simulations and dataset generation.}
  \label{fig:3D_designs}
\end{figure}

\begin{figure}[h!]
  \centering
  \begin{subfigure}[b]{0.24\textwidth}
    \centering
    \includegraphics[width=\textwidth]{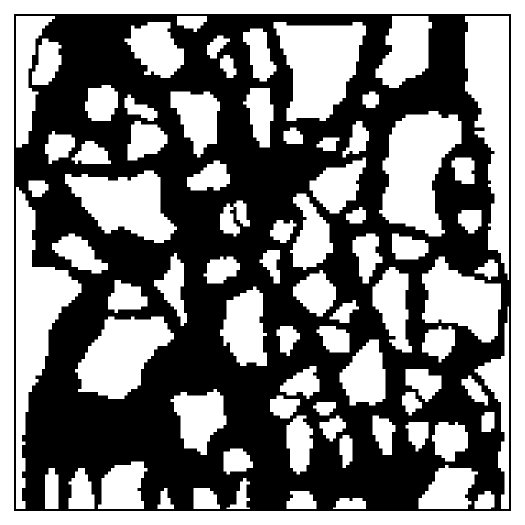}
    \caption{}
    \label{fig:filter2_3D}
  \end{subfigure}
  \hfill
  \begin{subfigure}[b]{0.24\textwidth}
    \centering
    \includegraphics[width=\textwidth]{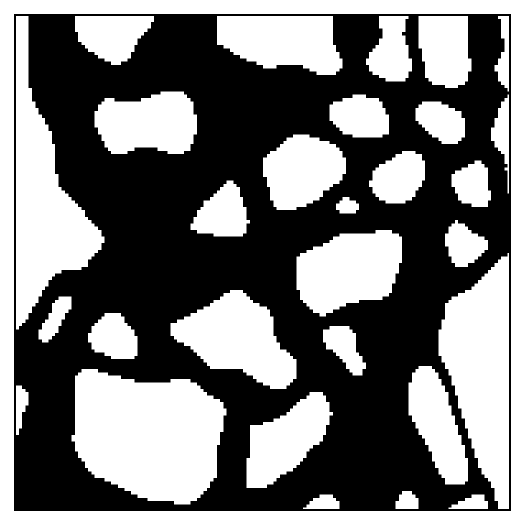}
    \caption{}
    \label{fig:filter6_3D}
  \end{subfigure}
  \hfill
  \begin{subfigure}[b]{0.24\textwidth}
    \centering
    \includegraphics[width=\textwidth]{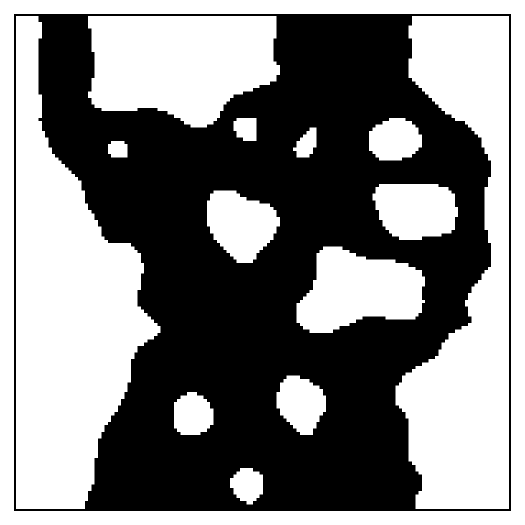}
    \caption{}
    \label{fig:filter8_3D}
  \end{subfigure}
  \hfill
  \begin{subfigure}[b]{0.24\textwidth}
    \centering
    \includegraphics[width=\textwidth]{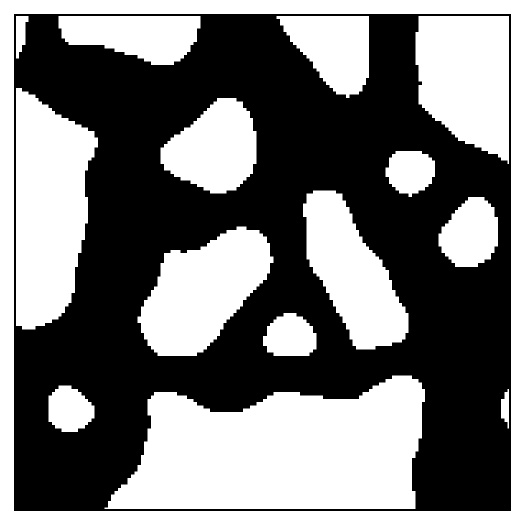}
    \caption{}
    \label{fig:filter10_3D}
  \end{subfigure}
  \caption{Optimized 3D designs obtained using different filter radii ($r = 
 2,6,8,10$ elements).}
\label{fig:filter_3d}
\end{figure}

Fig. \ref{fig:3D_designs} shows the optimized design corresponding to a filter radius of $r = 4$. Additional designs with varying filter sizes are provided in Fig. \ref{fig:filter_3d}. This selected design is used as a data generator: it undergoes a two-cycle tension–compression loading process to produce 3D strain–stress data. Using the resulting dataset, GRU models with three different initializations are trained and evaluated on a 3D random polynomial test set. The average performance of the three models is reported in Tab. \ref{tab:filter_radius_nrmse_3d}.

The results demonstrate that, even though the optimization explicitly targets only in-plane strain diversity, the GRU model still performs well on predicting the out-of-plane normal stress component $\sigma_{33}$. This suggests that promoting diversity in in-plane strain states also contributes to informative variation in out-of-plane responses. In contrast, the prediction accuracy for out-of-plane shear stresses is lower. This is attributed to the limited presence of such shear stress components in the training dataset generated from the thin specimen, resulting in insufficient exposure for the GRU to learn high-magnitude out-of-plane shear stress present in the test set.

\begin{table}[h!]
\caption{Prediction performance (NRMSE) of GRU models trained on strain–stress data generated from 3D optimized specimens with different filter radii.}
\label{tab:filter_radius_nrmse_3d}
\centering
\setlength{\tabcolsep}{0.30cm}
\renewcommand{\arraystretch}{1.5}
\newcolumntype{C}[1]{>{\centering\arraybackslash}p{#1}}
\begin{tabular}{C{2cm} C{1.5cm} C{1.5cm} C{1.5cm} C{1.5cm} C{1.5cm} C{1.5cm}}

\toprule
\multirow{2}{*}{\parbox[c]{1.5cm}{\centering \bfseries \shortstack{Filter \\[0.3em] Radius}}}
  & \multicolumn{6}{c}{\centering \bfseries NRMSE (\%)} \\
\cmidrule(l){2-7}
  & \bfseries $\sigma_{11}$ & \bfseries $\sigma_{22}$ & \bfseries $\sigma_{33}$  & \bfseries $\sigma_{12}$ & \bfseries $\sigma_{23}$ & \bfseries $\sigma_{13}$\\
\midrule
Filter\ 2  & 7.70  & 8.68  & 8.26  & 21.67 & 21.75 & 20.85\\ \midrule
Filter\ 4  & 12.97 & 13.18 & 13.00 & 32.77 & 36.92 & 36.26 \\ \midrule
Filter\ 6  & 14.30 & 14.71 & 15.05 & 40.12 & 39.27 & 40.47 \\ \midrule
Filter\ 8  & 16.75 & 19.22 & 19.85 & 55.79 & 54.02 & 53.42\\ \midrule
Filter\ 10 & 20.97 & 20.12 & 21.27 & 59.21 & 62.01 & 62.17\\ 
\bottomrule
\end{tabular}
\end{table}

\section{Supplementary Methods}

\subsection{Topology optimization setting}
\label{app: to_setting}

A square design domain with side length $L = 10\ \mathrm{mm}$ is selected whose bottom is fixed in the loading direction, and displacement loading 
$ u = 0.5\ \mathrm{mm}$ is applied on the top. The von Mises elastoplastic model with isotropic hardening is assumed to describe the material behavior. The material parameters are summarized in Tab.  \ref{tab:material_parameters}. 

\begin{table}[h!]
\caption{Material properties of the von Mises model with isotropic hardening.}
\label{tab:material_parameters}
\centering
\small
\setlength{\tabcolsep}{0.4cm}
\renewcommand{\arraystretch}{1.5}
\begin{tabular}{ccccc}
\toprule
\multicolumn{2}{c}{\bfseries Elastic} & \multicolumn{2}{c}{\bfseries Isotropic hardening} & \\ 
\cmidrule(lr){1-2}\cmidrule(lr){3-5}
$E$ (GPa) & $\nu$ & $Y_{0}$ (MPa) & $E_{t}$ (MPa) & \\ 
\midrule
110 & 0.33 & 900 & 500 & \\ 
\bottomrule
\end{tabular}

\vspace{0.2cm}
\begin{minipage}{0.9\linewidth}
\centering
\textbf{Isotropic hardening law:} $\sigma_{y}(\Bar{\varepsilon}^{p}) = Y_{0} + E_{t}\,\Bar{\varepsilon}^{p}$ \, (MPa)
\end{minipage}
\end{table}

The design domain is initialized with a uniform intermediate density of $ \rho_e = 0.5$. In addition, the elastic modulus of these intermediate (``gray") elements is interpolated based on the associated density, $\rho_e$, using the modified SIMP (Solid Isotropic Material with Penalization) scheme \cite{sigmund2007morphology}, e.g., 
\begin{equation}
E(\rho_e) = E_{\mathrm{min}} + \rho_e^p \left(E_0 - E_{\mathrm{min}}\right),
\end{equation}
where $E_0 = 110\,\mathrm{GPa}$ is the elastic modulus of the fully solid material, $E_{\mathrm{min}} = E_0 \times 10^{-9} = 110\,\mathrm{Pa}$ is the elastic modulus of the void material (a small nonzero value introduced to avoid singularity), and $p$ is the penalization exponent that drives the solution towards a discrete design. In this study, a continuously increased $p$ from $1$ to $10$ is applied. Given that we aim to maximize the strain diversity in the optimization process, the hardening law parameters in the plasticity model are kept constant and not interpolated to avoid the premature yielding of the "gray" elements. In the end, the continuous density field is thresholded to produce a binary material distribution. During the optimization process, a volume fraction constraint is imposed such that the volume fraction is within the range of [40\%, 60\%]. 

\subsection{Filter in topology optimization procedure}

\label{app:cone_filter}
A conventional spatial filter is used to avoid checkerboard patterns arising in density-based topology optimization \cite{bourdin_filters_2001}. This regularizes the solution and also imposes a prescribed minimum length scale, leading to the following filtered density at each element $e$:

\begin{equation}
\tilde{\rho}_e = \frac{\sum_{i \in \mathcal{N}_e} w(\mathbf{x}_i) \rho_i}{\sum_{i \in\mathcal{N}_e} w(\mathbf{x}_i)},
\label{eq:cone}
\end{equation}
where $\mathcal{N}_e$ is the neighborhood set of element $e$, $x_i$ is the centroid of the element $i$, and the weight function is defined by a cone-shaped function:
\begin{equation}
w(\mathbf{x}_i) = r_{\mathrm{min}} - \|\mathbf{x}_i - \mathbf{x}_e\|,
\label{eq:weight_function}
\end{equation}

\noindent where $r_{\mathrm{min}}$ is the prescribed filter radius. However, note that the filtering operation still leaves elements with intermediate densities. Enforcing a near-binary material distribution becomes possible by further processing the density field with the smooth threshold projection \cite{xu_volume_2010},

\begin{equation}
\hat{\rho}_i\!\left(\tilde{\rho}_i;\,\beta,\eta\right)
  =\frac{\tanh\!\bigl(\beta\,\eta\bigr)
         +\tanh\!\bigl[\beta\,\bigl(\tilde{\rho}_i-\eta\bigr)\bigr]}
        {\tanh\!\bigl(\beta\,\eta\bigr)
         +\tanh\!\bigl[\beta\,\bigl(1-\eta\bigr)\bigr]} ,
\label{eq:xu_projection}
\end{equation}
 
\noindent where $\eta$ is the threshold parameter that controls the switch from 0 to 1 (fixed to $0.5$), and $\beta$ controls the steepness of the projection by continuously increasing from $1$ to $10$ as the iterations progress. This continuation technique incrementally refines material–void transitions while maintaining numerical stability in the initial rounds, eventually guiding the design towards an optimized topology that is nearly binary (each element can be fabricated with a single material).
\subsection{Generating 3D random polynomial test set}
\label{app:3D_testset}

\begin{figure}[h!]
	\centering
	\begin{subfigure}[b]{0.40\textwidth}
		\centering
		\includegraphics[width=\textwidth]{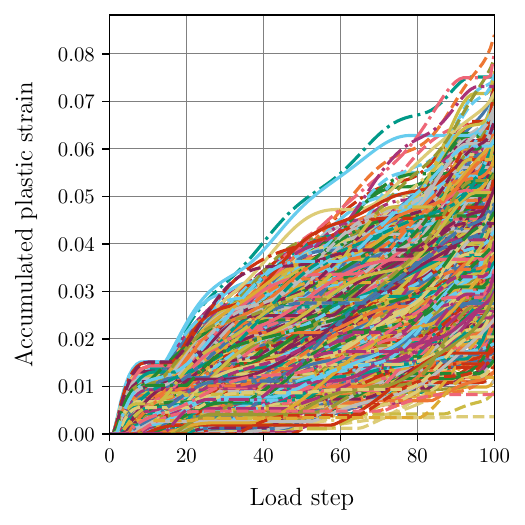}
		\caption{}
	\end{subfigure}%
	\hfill
	\begin{subfigure}[b]{0.40\textwidth}
		\centering
		\includegraphics[width=\textwidth]{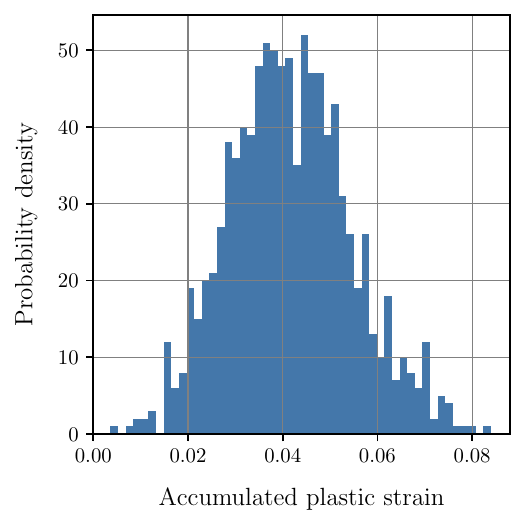}
		\caption{}
	\end{subfigure}
	\hfill
	\begin{subfigure}[b]{0.40\textwidth}
		\centering
		\includegraphics[width=\textwidth]{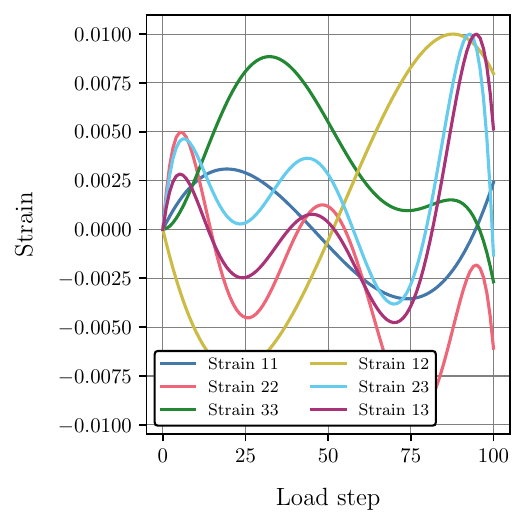}
		\caption{}
	\end{subfigure}
	\hfill
	\begin{subfigure}[b]{0.40\textwidth}
		\centering
		\includegraphics[width=\textwidth]{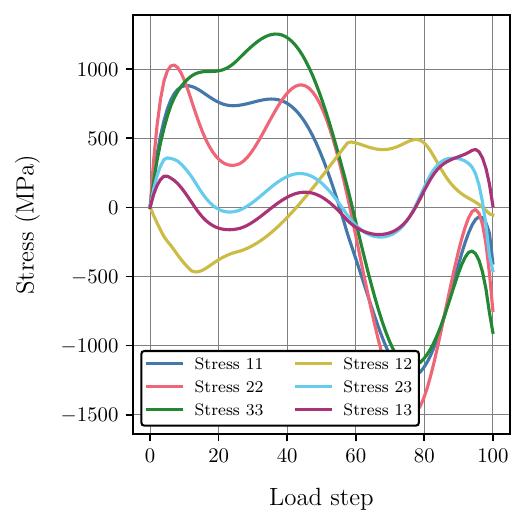}
		\caption{}
	\end{subfigure}
	\caption{Visualization of the 3D test set: 
		(a) Accumulated plastic strain versus load step; 
		(b) Probability density distribution of the accumulated plastic strain across the test set; 
		(c) A representative sample of strain paths; 
		(d) The corresponding stress response for the strain paths shown in (c).}
	
	\label{fig:3D_testset_sample}
\end{figure}

The 3D test set is generated using the same procedure as the 2D case, with the key difference being a narrower strain range. Specifically, each of the six strain components is sampled within the bounds of [$-1\%$, $1\%$]. These tighter bounds are selected to prevent excessively large stress magnitudes during evaluation.
An example testing path is shown in Fig. \ref{fig:3D_testset_sample}.

\textit{Remark:} Because the 3D test set is generated from random combinations of six independent strain components, and given the open yield surface characteristic of the 3D von Mises plasticity model, larger strain bounds can result in unrealistically large stress values. Such stresses are physically unrealistic and undesirable for reliable model evaluation.

\begin{figure}[h!]
	\centering
	\begin{subfigure}[b]{0.30\textwidth}
		\centering
		\includegraphics[width=\textwidth]{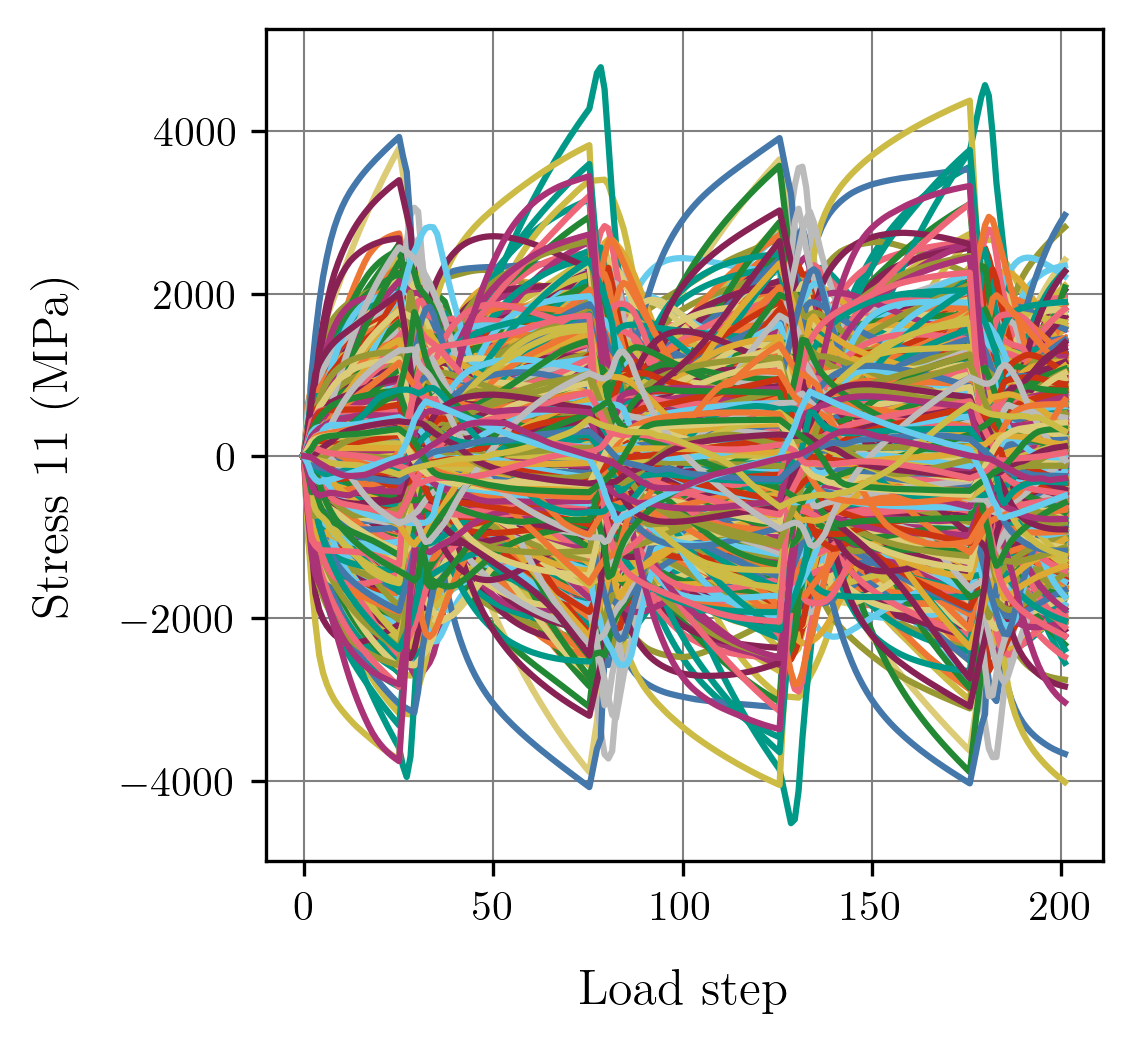}
		\caption{}
	\end{subfigure}%
	\hfill
	\begin{subfigure}[b]{0.30\textwidth}
		\centering
		\includegraphics[width=\textwidth]{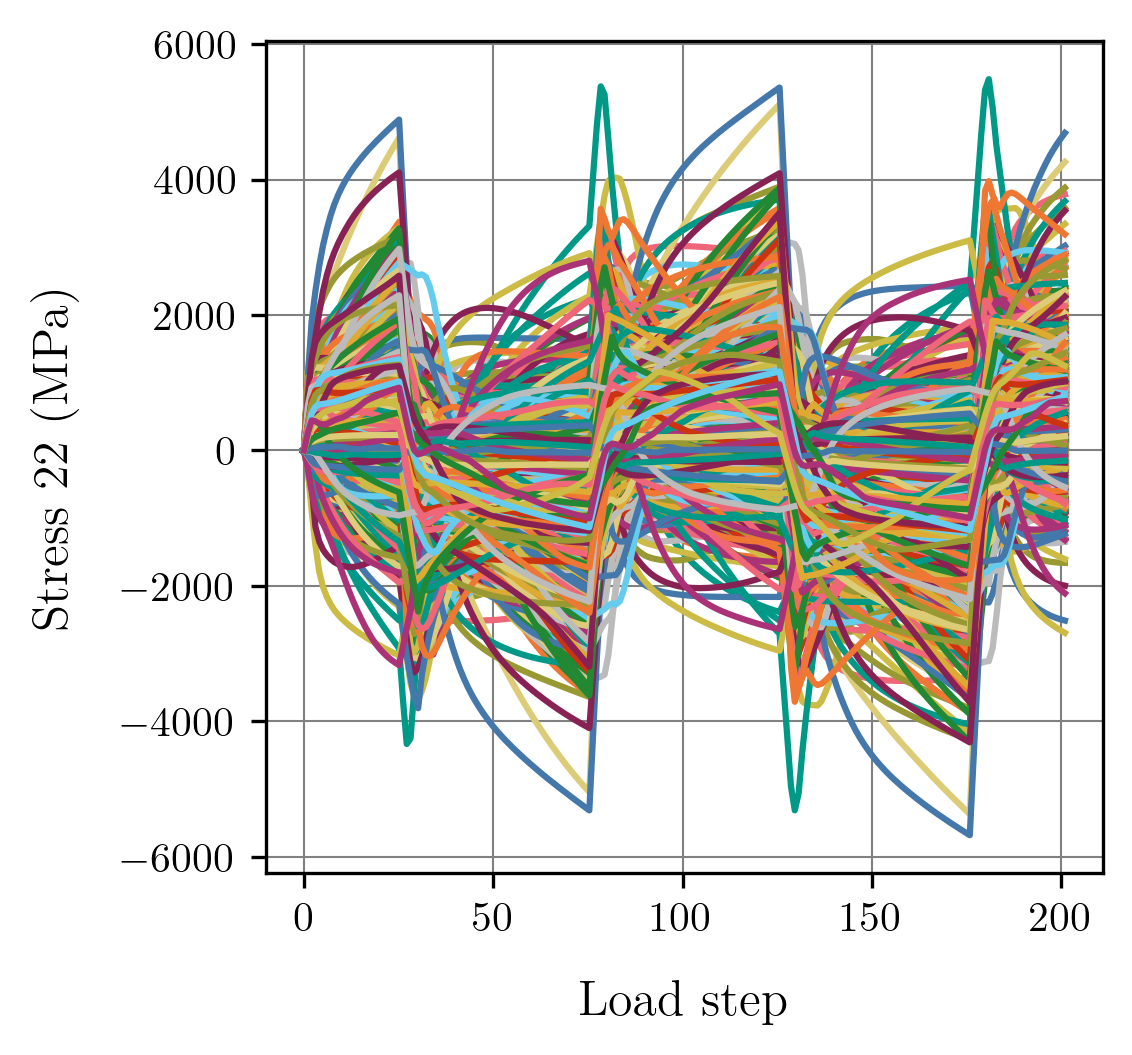}
		\caption{}
	\end{subfigure}%
	\hfill
	\begin{subfigure}[b]{0.30\textwidth}
		\centering
		\includegraphics[width=\textwidth]{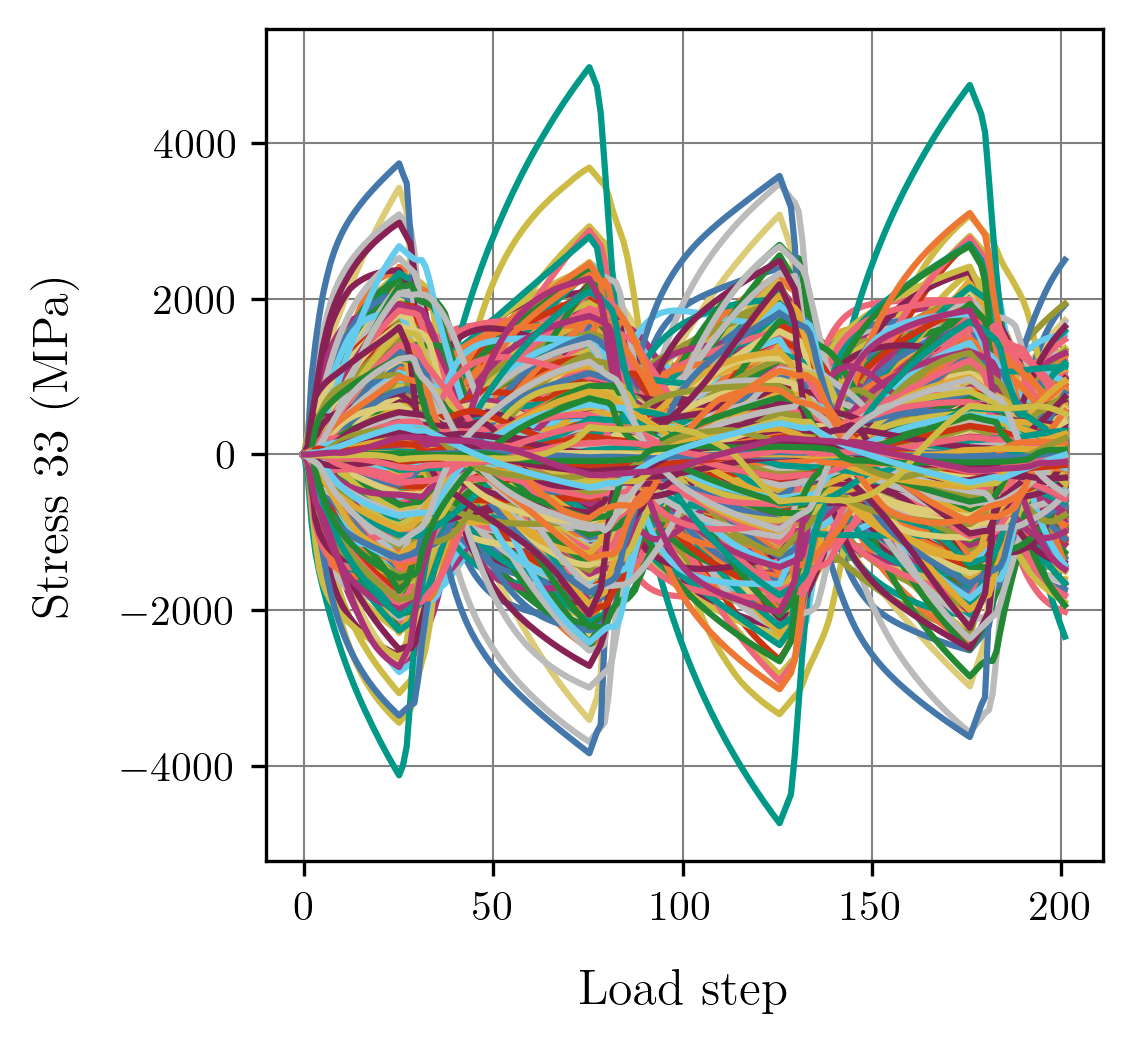}
		\caption{}
	\end{subfigure}
	\begin{subfigure}[b]{0.30\textwidth}
		\centering
		\includegraphics[width=\textwidth]{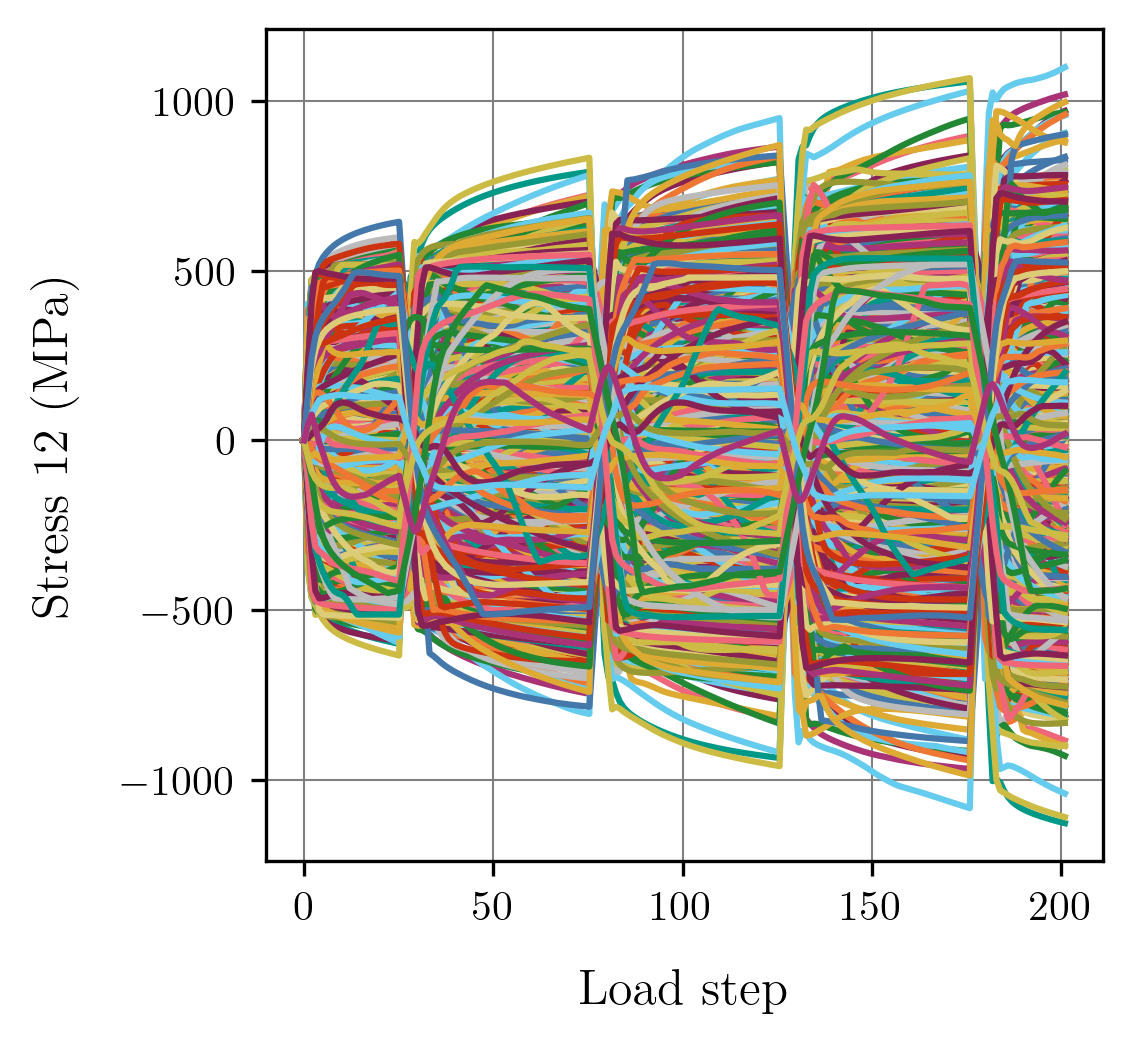}
		\caption{}
	\end{subfigure}%
	\hfill
	\begin{subfigure}[b]{0.30\textwidth}
		\centering
		\includegraphics[width=\textwidth]{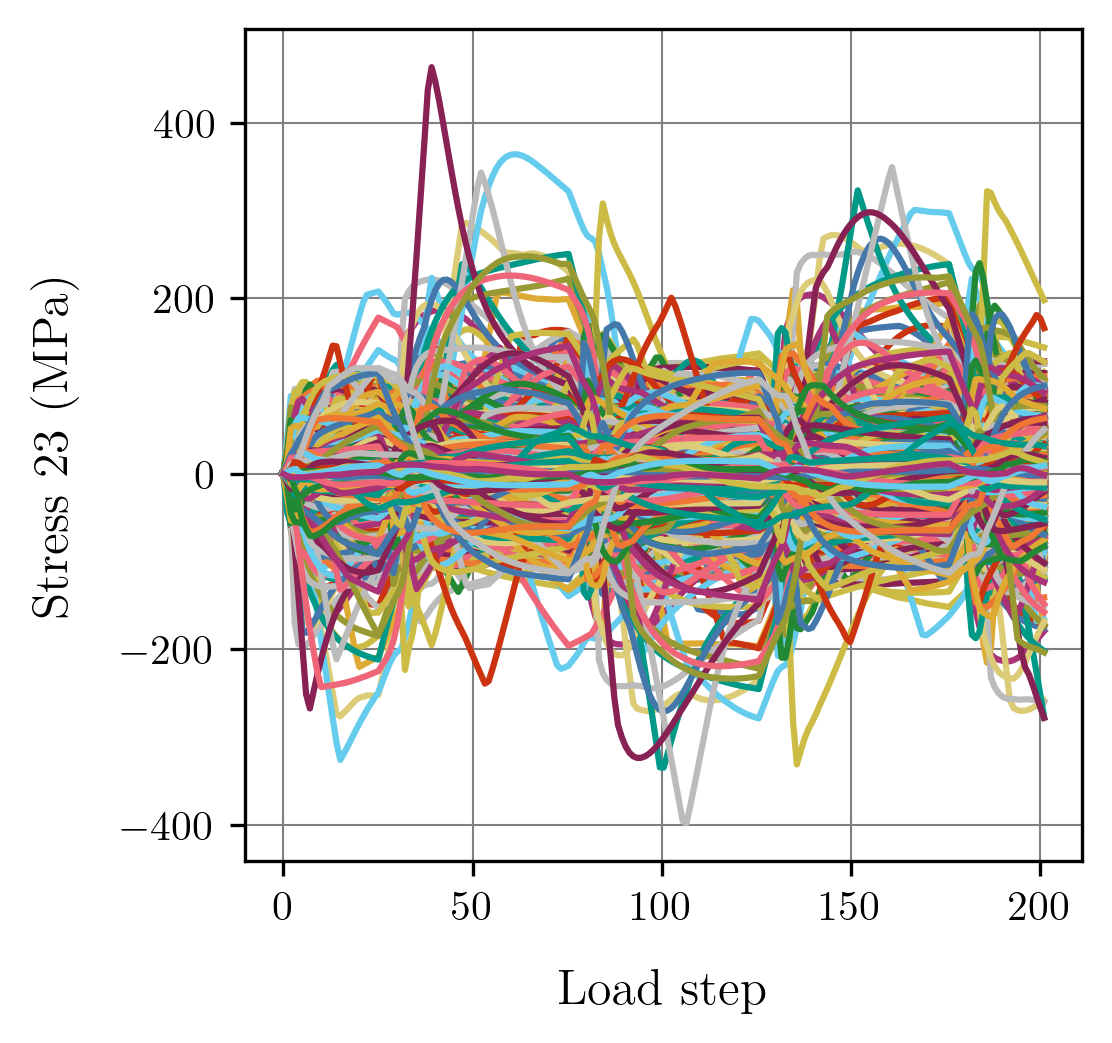}
		\caption{}
	\end{subfigure}%
	\hfill
	\begin{subfigure}[b]{0.30\textwidth}
		\centering
		\includegraphics[width=\textwidth]{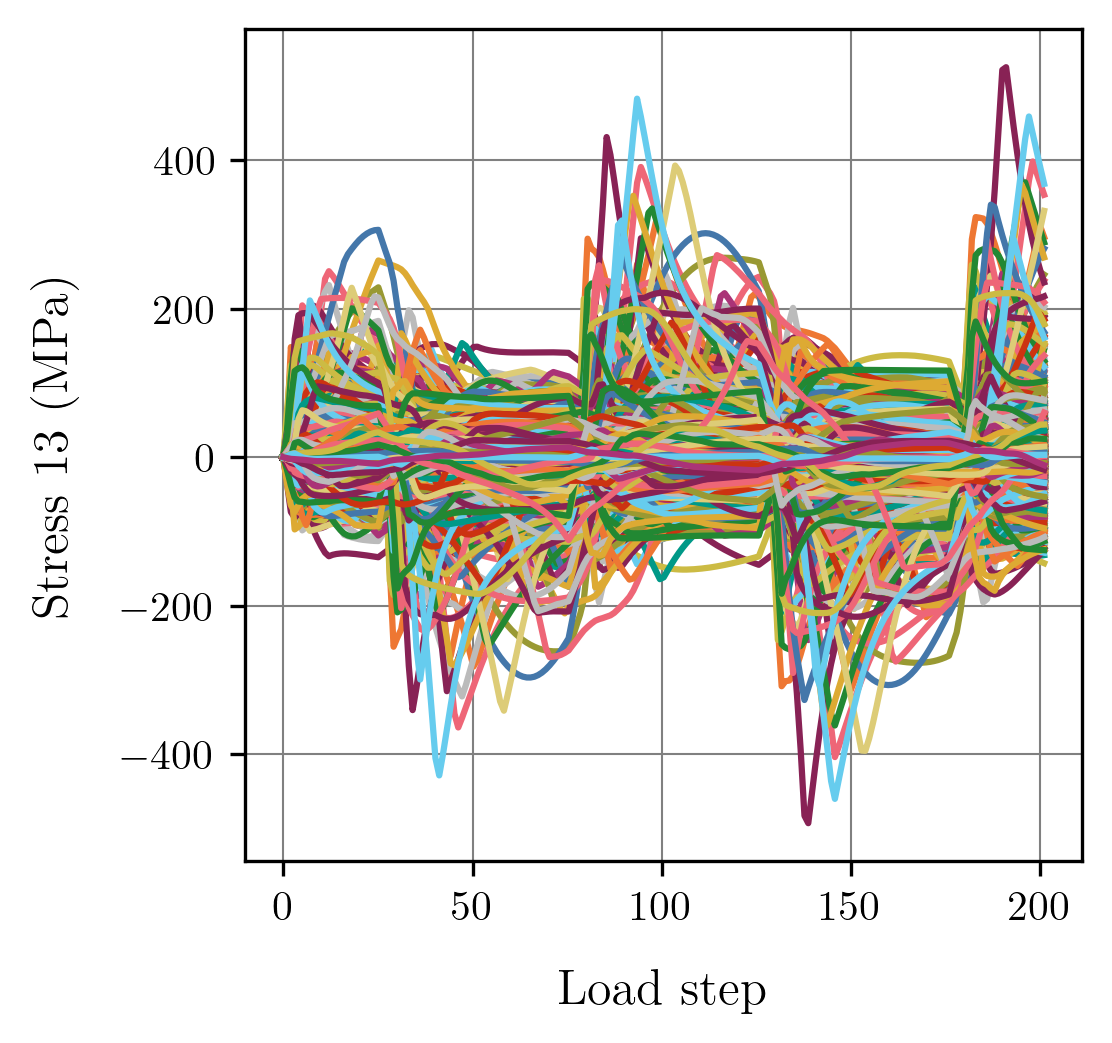}
		\caption{}
	\end{subfigure}
	\caption{Stress history of 3D-optimized specimen dataset.}
	\label{fig:stress_optimized3d}
\end{figure}

\begin{figure}[h!]
  \centering
  \begin{subfigure}[b]{0.30\textwidth}
    \centering
    \includegraphics[width=\textwidth]{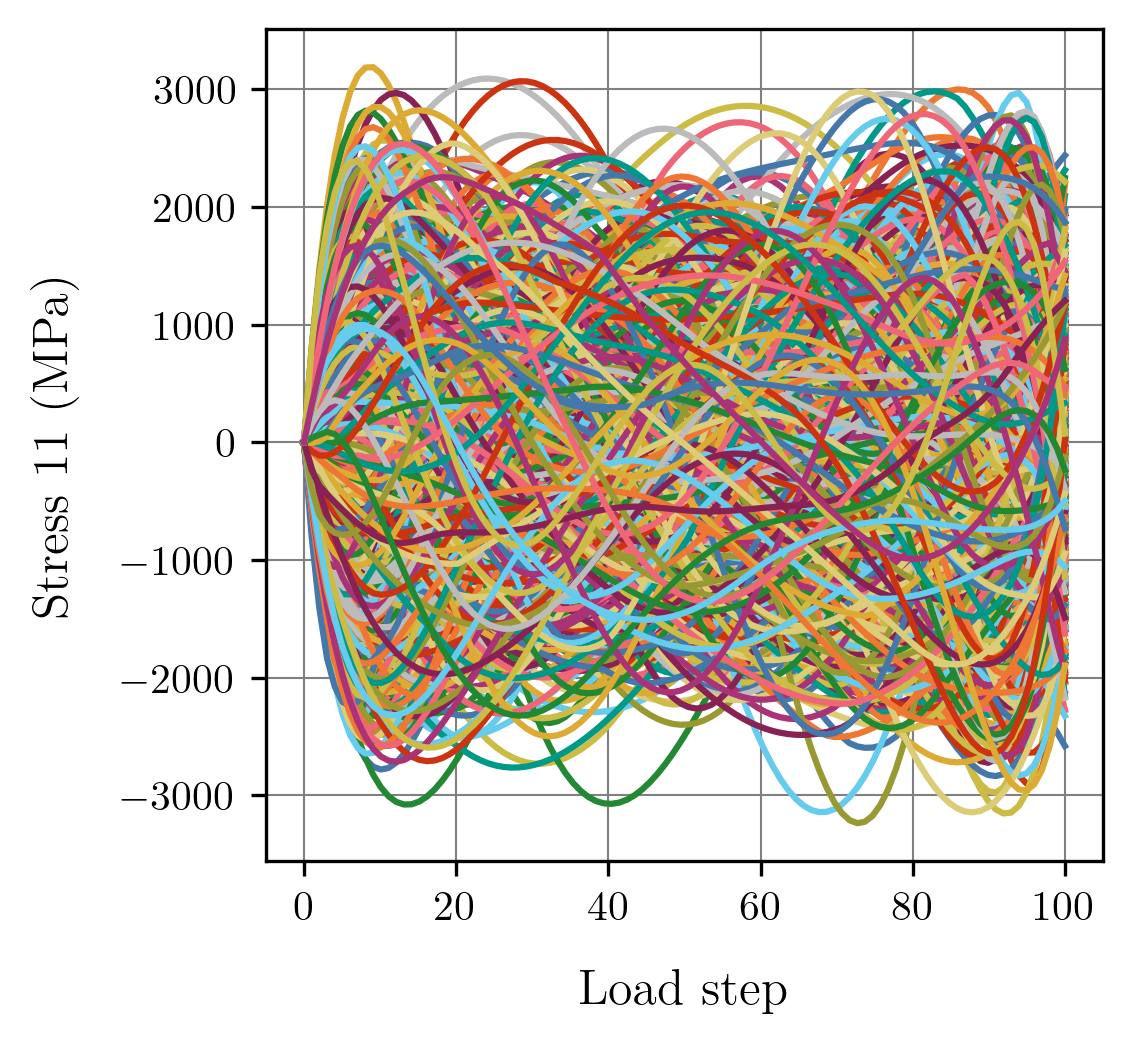}
    \caption{}
  \end{subfigure}%
  \hfill
  \begin{subfigure}[b]{0.30\textwidth}
    \centering
    \includegraphics[width=\textwidth]{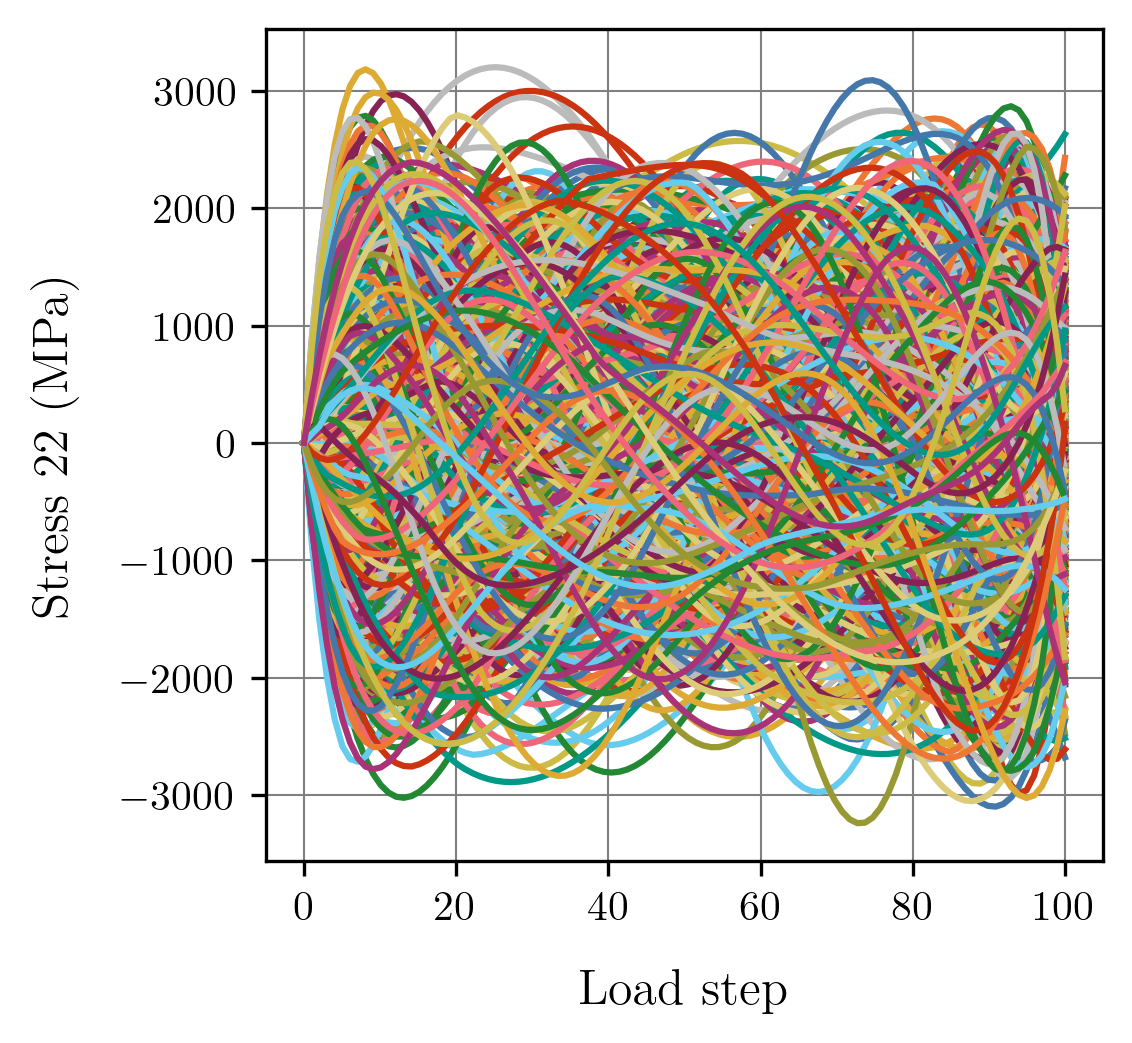}
    \caption{}
  \end{subfigure}%
  \hfill
  \begin{subfigure}[b]{0.30\textwidth}
    \centering
    \includegraphics[width=\textwidth]{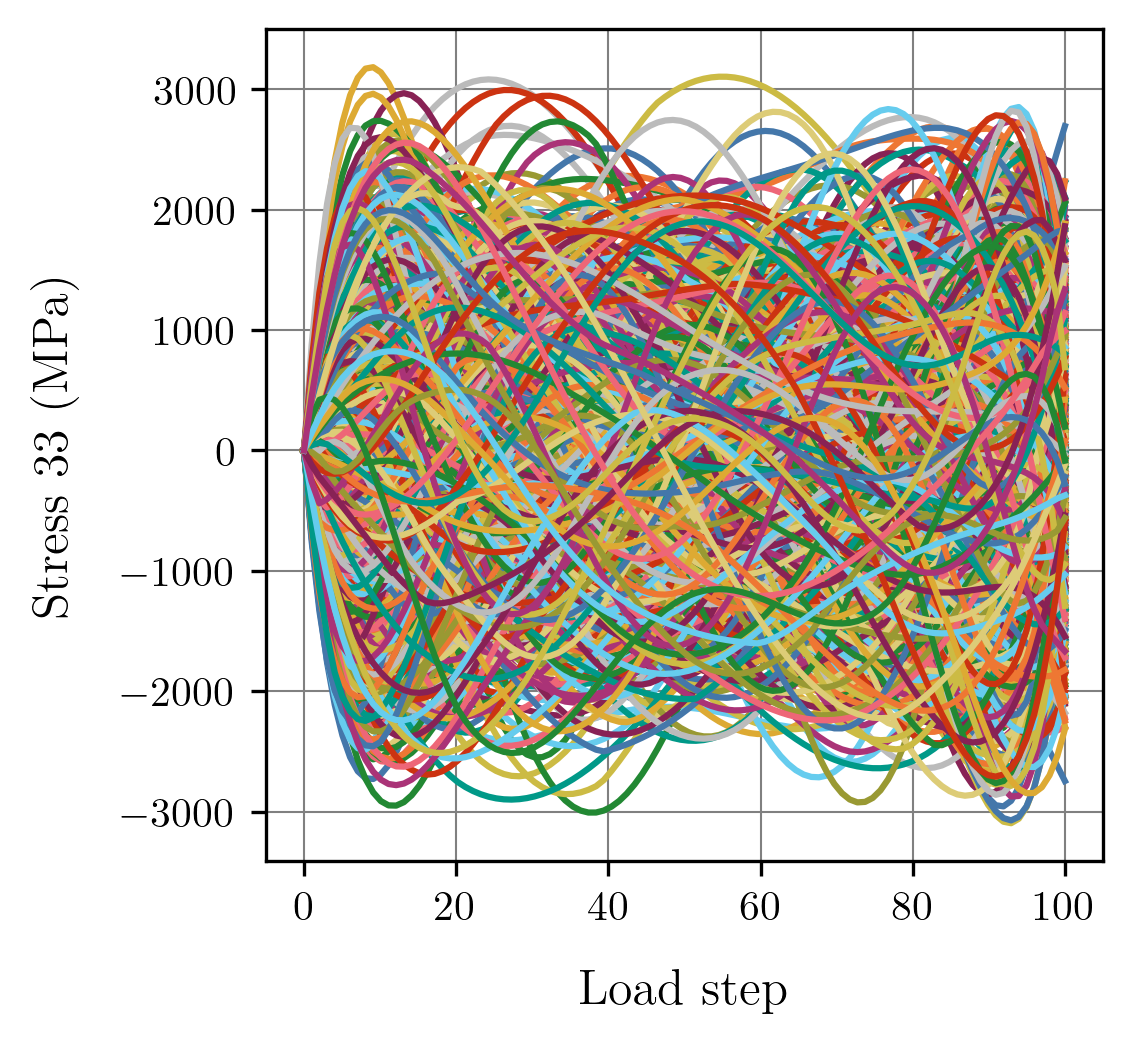}
    \caption{}
  \end{subfigure}
    \begin{subfigure}[b]{0.30\textwidth}
    \centering
    \includegraphics[width=\textwidth]{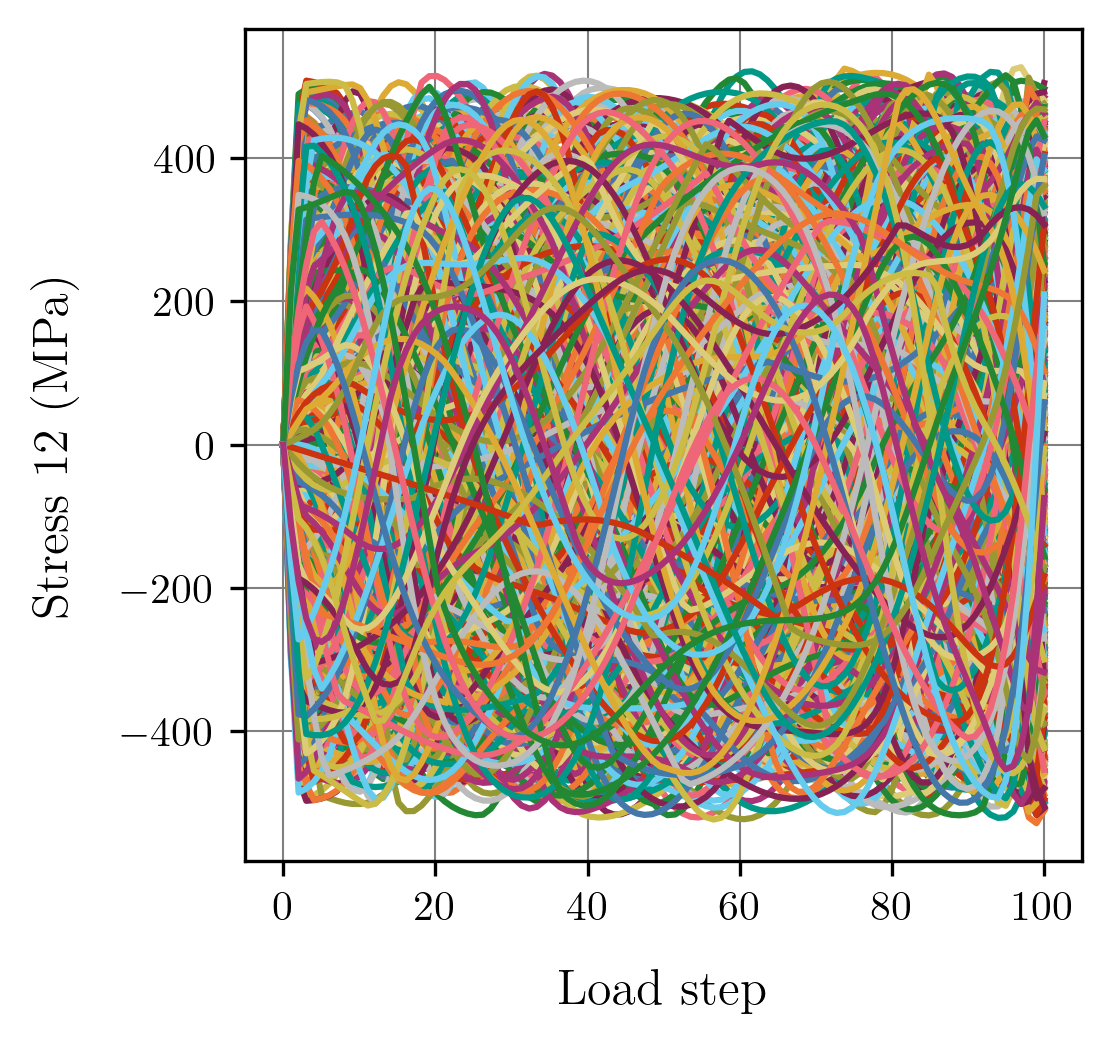}
    \caption{}
  \end{subfigure}%
  \hfill
  \begin{subfigure}[b]{0.30\textwidth}
    \centering
    \includegraphics[width=\textwidth]{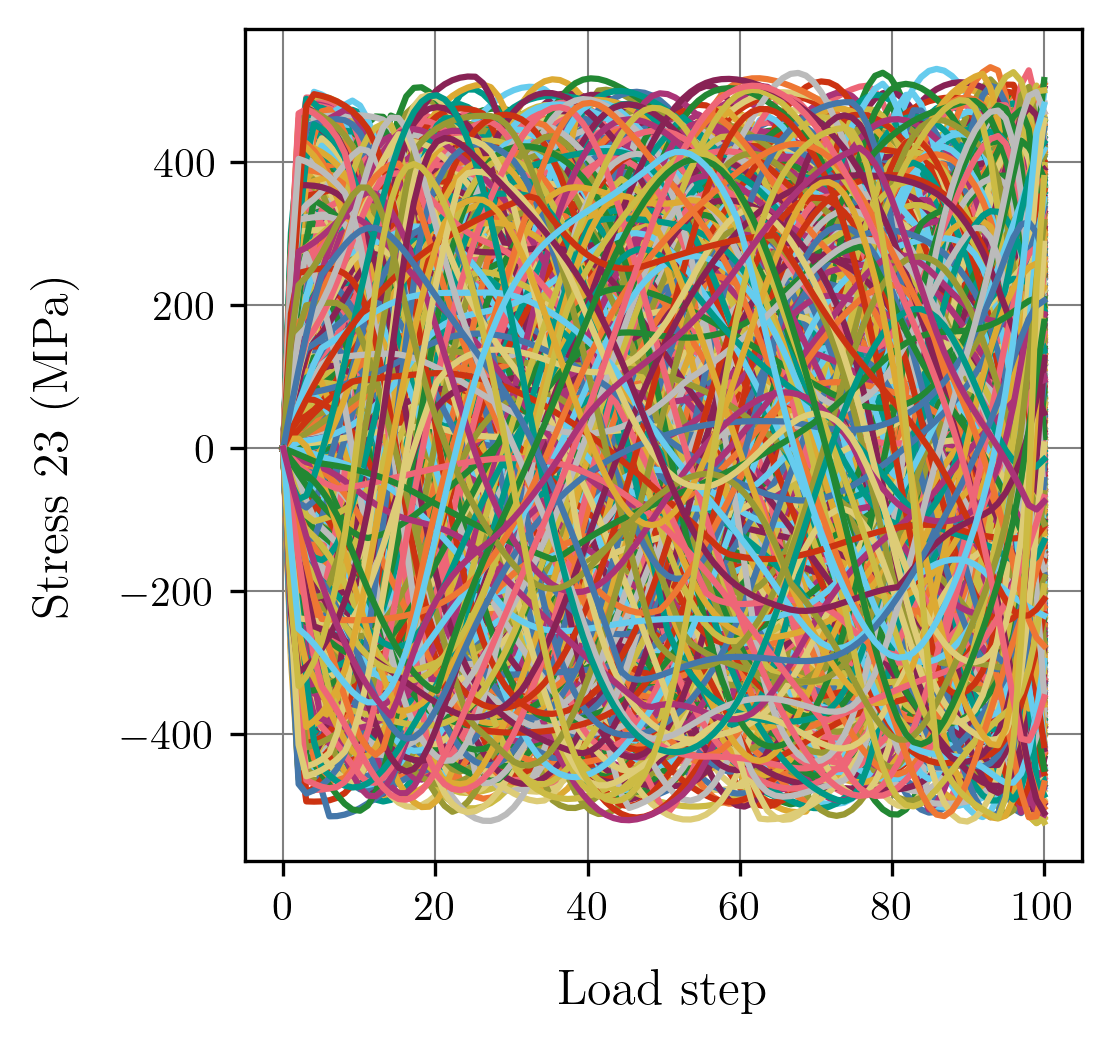}
    \caption{}
  \end{subfigure}%
  \hfill
  \begin{subfigure}[b]{0.30\textwidth}
    \centering
    \includegraphics[width=\textwidth]{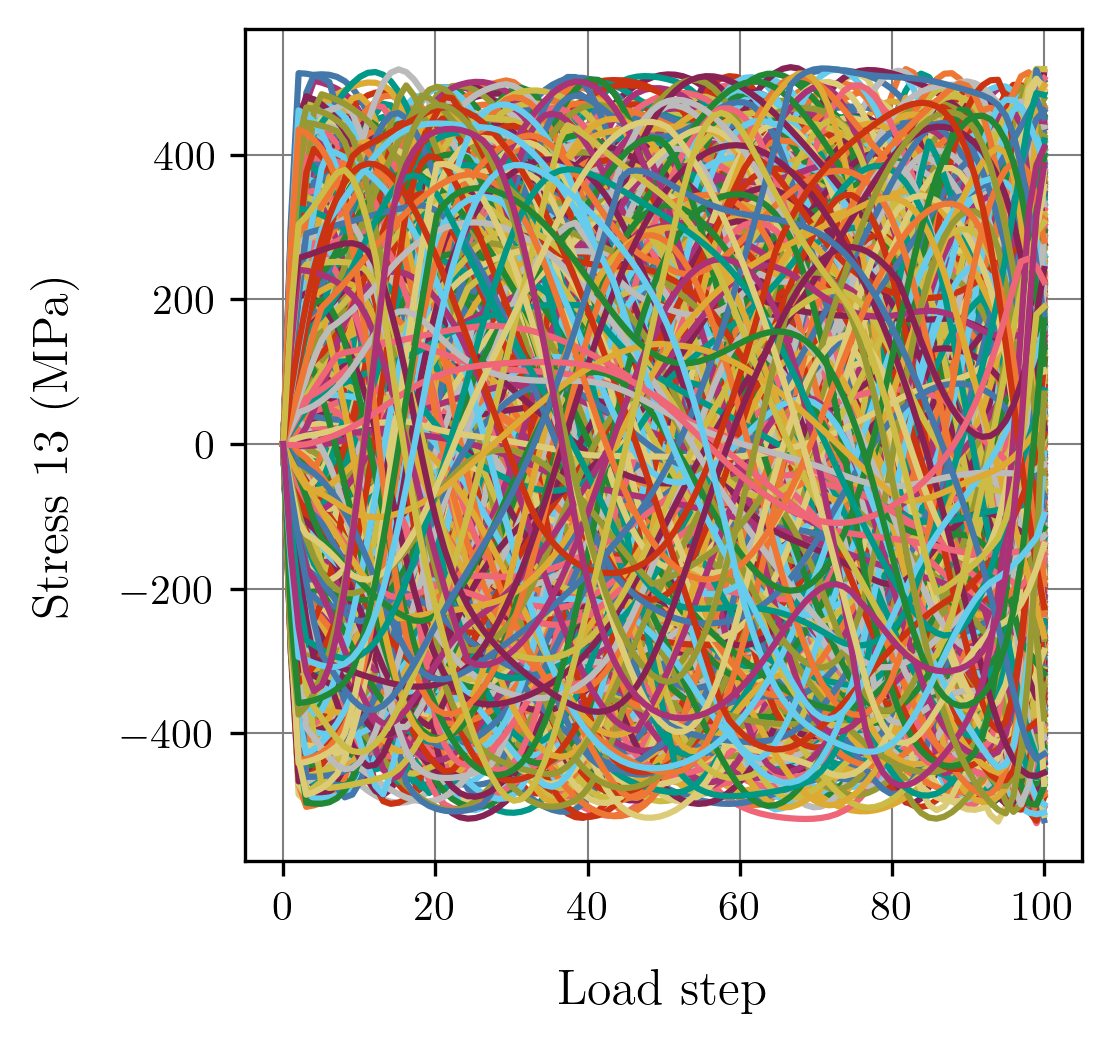}
    \caption{}
  \end{subfigure}
  \caption{Stress history of 3D testset. }
  \label{fig:stress_testset3d}
\end{figure}

For comparison, the stress histories obtained from the optimized specimen and from the testing set are presented in Fig. \ref{fig:stress_optimized3d} and Fig. \ref{fig:stress_testset3d}, demonstrating that the stress ranges in both datasets are comparable.

\clearpage

\bibliographystyle{elsarticle-num}
\bibliography{reference}

\end{document}